\documentclass[10pt, aps, prd,
reprint, 
notitlepage, nofootinbib, longbibliography,
superscriptaddress,
]{revtex4-1}

\usepackage{amsmath,bm,amsfonts}
\usepackage{pstricks,pstricks-add}
\usepackage{pst-all,pst-slpe}
\usepackage{graphicx}
\usepackage[breaklinks=true,colorlinks=true,urlcolor=blue,linkcolor=red,
    citecolor=red]{hyperref}

\begin{document}

\title{Geometrical dependence in Casimir-Polder repulsion: Anisotropically polarizable
atom and anisotropically polarizable annular dielectric}

\author{John Joseph Marchetta}
\email{jjmarchetta@gmail.com} 
\affiliation{Department of Physics,
Southern Illinois University--Carbondale, Carbondale, Illinois 62901, USA}

\author{Prachi Parashar}
\email{Prachi.Parashar@jalc.edu} 
\affiliation{John A. Logan College, Carterville, Illinois 62918, USA}

\author{K. V. Shajesh}
\email{kvshajesh@gmail.com}
\affiliation{Department of Physics, 
Southern Illinois University--Carbondale, Carbondale, Illinois 62901, USA}

\date{\today}
\begin{abstract}

Casimir-Polder interaction energies between a point anisotropically
polarizable atom and an annular dielectric are shown to exhibit
localized repulsive long-range forces in specific configurations.
We show that when the atom is positioned at the center of the annular
dielectric, it is energetically favorable for the atom to align its
polarizability with respect to that of the dielectric. As the atom
moves away from, but along the symmetry axis of the annular dielectric,
it encounters a point where the polarizable atom experiences no torque
and the energy is free of orientation dependence. At this height,
abruptly, the atom prefers to orient its polarizability perpendicular
to that of the dielectric. For certain configurations, it encounters
another torsion-free point a larger distance away, beyond which it
prefers to again point its polarizability with respect to that of the
dielectric. We find when the atom is close enough, and oriented such
that the energy is close to maximum, the atom could be repelled. For
certain annular polarizations, repulsion can happen below or above
the torsion-free height. Qualitative features differ when the atom is
interacting with a ring, versus a plate of infinite extent with a
hole. In particular, the atom can prefer to orient perpendicular to the
polarizability of the plate at large distances, in striking contrast
to the expectation that it will orient parallel. To gain insight of
this discrepancy, we investigate an annular disc, which captures the
results of both geometries in limiting cases. These energies are too
weak for immediate applications, nevertheless, we elaborate an
interesting application on a prototype of a Casimir machine using
these configurations.

\end{abstract}

\maketitle

\section{Introduction}

\begin{figure}
\includegraphics[width=8cm]{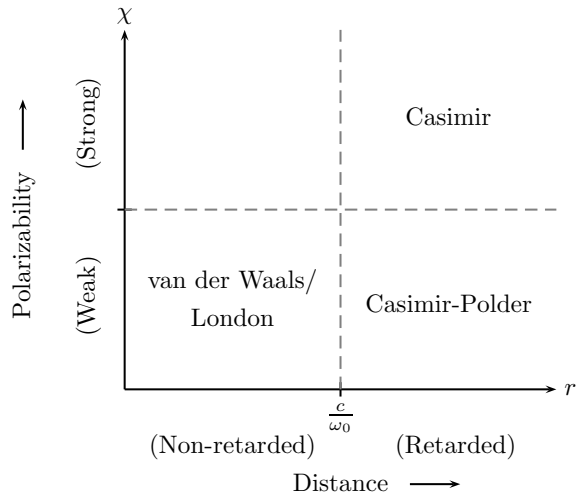}
\caption{Diagram showing terminology associated with
interactions mediated by quantum fluctuations in electrodynamics.
Here $\chi$ is the dielectric susceptibility of a material
characterized by plasma frequency $\omega_p$,
$\omega_0$ is a characteristic frequency of the material,
$c$ is speed of light in vacuum, and $r$ is
the distance between two objects. }
\label{term-def-fig}
\end{figure}

The term Casimir effect has come to refer to the entire phenomena 
associated with quantum fluctuations in electrodynamics. 
Figure~\ref{term-def-fig} shows a schematic diagram
that identifies the various domains in interactions mediated
by quantum fluctuations. 
The van der Waals~\cite{Waals:1873sl} and London dispersion 
forces~\cite{London:1930a,London:1930b,Hettema:2001cq}
govern these interactions in the non-retarded weak regime,
while the Casimir-Polder interaction energy~\cite{Casmir:1947hx}
involves interactions between weak dielectric materials in the
retarded regime. The energies in the weak regime
are often easier to compute. In contrast, the
Casimir energy between highly conducting materials
in the retarded regime are typically hard to compute
because it involves identification and subtraction of divergences
in non-trivial ways to arrive at the expression for energy.
The Casimir energy is also very sought-after because it is a
manifestation of zero point energy associated to the quantum 
vacuum~\cite{Casimir:1948pc, Milton:2011fpc}.
The limiting forms in Fig.~\ref{term-def-fig} are generalized
by the Lifshitz formula~\cite{Lifshitz:1956sb},
and were further generalized
and studied by Dzyaloshinskii, Lifshitz,
and Pitaevskii (DLP)~\cite{Dzyaloshinskii:1961fw}.

The Casimir effect most often
leads to attractive forces between objects, and
the plausibility of associated repulsive forces has always
captivated attention.
In the non-retarded van der Waals regime of Fig.~\ref{term-def-fig},
repulsion has been predicted in three-body interactions
at least since the work of
Axilrod and Teller~\cite{Axilrod:1943at}
and Muto \cite{Muto:1943fc}. In the
Casimir-Polder regime, repulsion between anisotropically
polarizable atoms was predicted in the works of
Craig and Power~\cite{Craig1969ma,Craig:1969pa} and 
has been extensively explored since then;
see Ref.~\cite{Babb2005ac} and references therein. However,
it was not until a decade ago when it was realized that
similar repulsive behavior was plausible in the
Casimir regime of Fig.~\ref{term-def-fig}.
This came by in Ref.~\cite{Levin:2010vo}, where it was 
argued on physical grounds that the interaction energy 
between an elongated needle-shaped-conductor and a perfectly conducting 
metal sheet with a circular aperture could have a local minima.
They showed that the Casimir force could become repulsive
when the needle got sufficiently close to the aperture.
This also meant that the force between a charge placed in
front a perfectly conducting object could be
repulsive~\cite{Levin:2011a}, and similar configurations were
further explored in Ref.~\cite{McCauley:2011rd}. 
It was realized there that the anisotropy in the geometry of a highly
conducting object corresponds to an effective anisotropic permittivity
and the interplay of these anisotropic permittivities
leads to non-monotonic interaction energies that show repulsion. 
These remarkable predictions were based on symmetry arguments
and were confirmed using numerical calculations.

Here we list some of the attempts made towards understanding
the result in Ref.~\cite{Levin:2010vo} analytically.
In Ref.\cite{Eberlein2011hwp}, it was shown that even
in the non-retarded van der Waals regime,
similar configurations lead to repulsion.
They used the method of inversion that was successfully
used in the heyday of electrostatics. 
In Ref.~\cite{Abrantes2018tcp}, the method of inversion was
again used to show that in the non-retarded van der Waals regime
an anisotropically polarizable atom placed along the 
symmetry axis of a toroid could experience repulsion. 
In Refs.~\cite{Shajesh:2011daa} and \cite{Milton2012esc},
it was shown that in the retarded Casimir-Polder regime
of Fig.~\ref{term-def-fig}, repulsion is possible in
configurations with anisotropically polarizable atoms
and anisotropic dielectric materials.
An analytic formula for the Casimir-Polder energy between
dielectric bodies was derived in 
Refs.~\cite{Shajesh:2011daa} and \cite{Milton2012esc}.
In particular, a closed form expression for the force between an
anisotropically polarizable atom and a dielectric plate with a circular
aperture was shown to demonstrate repulsion.
In spite of the above successes in the weak scenario,
and having numerically found configurations exhibiting repulsion
in the Casimir regime of Fig.~\ref{term-def-fig},
analytic derivation of a closed-form expression for
the interaction energy between a polarizable atom
and a highly conducting plate
with an aperture still remains an open problem. 
Initiative towards such a derivation has been presented in
Refs.~\cite{Milton2011asa} and \cite{2012:Miltonfpc}. 
Another proposal was to exploit the axial symmetry of the
configuration~\cite{Shajesh2017ssa}.
However, a closed-form analytic result remains elusive.

In this paper we investigate the interaction energy between two
neutral anisotropically polarizable objects mediated
by quantum fluctuations. 
We ignore magnetization effects here.
The outline of the paper
and summary of results are as follows:
In Sec.~\ref{electrostatics-sec},
we derive the interaction energy between a point permanent dipole
and an electrically polarized ring.
We give a summary of the formalism in Sec.~\ref{sec-formApp},
and present a few more configurations related to annular plate
in Sec.~\ref{annular-plate-section}.
We take up the study of a point anisotropically polarizable atom
and an anisotropically polarizable dielectric ring using
methods of Ref.~\cite{Shajesh:2011daa} in Sec.~\ref{sec-atom-ring}.
We show that for specific orientations
of the polarizability, the atom experiences a repulsive force
when the atom is very close to the ring.
We analyze the orientation dependence and distance dependence
of the interaction energies.
We also observe that for specific orientations the atom experiences a
second repulsive region along the symmetry axis, which is 
a new finding here.
In Sec.~\ref{annular-disc-sec}, we derive the interaction energy
between an atom and a dielectric annular disc. For large outer
radius, this energy approaches the interaction energy between an
atom and a plate with circular aperture. When the outer radius
and inner radius of the annular disc are close to each other, the
energy approaches that of a ring.
In Sec.~\ref{sec-summary-disc}, we highlight and summarize our results.
We point out that there exists two torsion free points on each side
of the annular disc, where the interaction energy is orientation
independent. The position of these points vary with geometric changes
to the disk, and differently according to the polarizability;
determining large distance orientation dependence.
In the ring limit of the annular disc, new repulsion emerges.
A brief summary of our results here has been separately
presented in Ref.~\cite{Marchetta:2020sap}, and the exposition
here serves as supplementary material for the article.
In Sec.~\ref{cas-machine-sec}, we describe an application of
these results in the construction of a Casimir machine.
We give concluding remarks in Sec.~\ref{conclusion-sec}. 

For completeness, we list two other classes of repulsive behaviors
that are possible between neutral polarizable objects.
One of these was mentioned in Ref.~\cite{Dzyaloshinskii:1961fw}
in which repulsion is possible whenever the dielectric
strength of the intervening medium is stronger relative
to the surrounding media.
The second is associated to repulsion arising from an
interplay of electric and magnetic properties of the
neutral objects~\cite{Feinberg1968pw}.
In the current level of our understanding, apparently,
these repulsions have different origins from those discussed
in this paper.
The repulsion studied in this article is geometric in origin
due to the presumable direct relationship between
the effective anisotropies in the dielectric susceptibility and
the respective anisotropies in the geometric shape of the conductor.
The repulsive nature of self interaction of a single conducting
sphere~\cite{Boyer:1968uf} could also probably be related
to this geometric origin, however it is not
an interaction energy between two objects as it is for
cases we are considering here.

\section{Electrostatic analog:
Point electric dipole above an electrically polarized ring}
\label{electrostatics-sec}

Let us begin with an example involving neutral objects having
permanent electric polarizations.
It will be a prelude for the following discussion
involving only neutral induced dipoles.

\begin{figure}
\includegraphics[width=8.5cm]{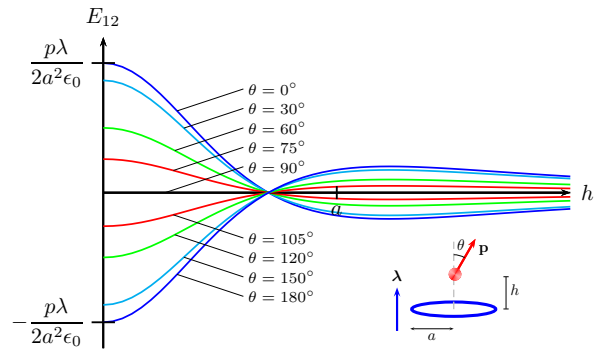}
\caption{
Electrostatic interaction energy between a point electric dipole ${\bf p}$ 
and a polarized ring of polarization ${\bm\lambda}$ as given by
Eq.\,(\ref{edapr}). Here $E_{12}$ is plotted with respect to $h$
for $\theta=0^\circ, 30^\circ, 60^\circ, 75^\circ$. The energy at
$\theta=0$ and $h=0$ is $p\lambda 2\pi/a^2 4\pi\varepsilon_0$. 
}
\label{atom-above-dielectric-ring-fig}
\end{figure}

A point (permanent) electric dipole ${\bf p}$ at position ${\bf r}_0$
is suitably described by the charge density
\begin{equation}
\rho_1({\bf x}) 
= - {\bf p} \cdot {\bm\nabla} \delta^{(3)}({\bf x}-{\bf r}_0),
\label{cdenppd}
\end{equation}
and a (permanent) electrically polarized ring of radius $a$
with dipole per unit length or polarization ${\bm\lambda}$
is described by the charge density
\begin{equation}
\rho_2({\bf x}^\prime) 
= -{\bm\lambda} \cdot {\bm\nabla}^\prime 
\delta(z^\prime-z_a) \delta(\rho^\prime-a),
\label{cdenprd}
\end{equation}
such that the plane of the ring is perpendicular to the
$z$ axis at $z^\prime=z_a$ with center of the ring on the $z$ axis,
see inset in Fig.~\ref{atom-above-dielectric-ring-fig}.
The electrostatic interaction energy between the point dipole
and the ring, in SI units, is given by
\begin{equation}
E_{12} = \frac{1}{4\pi\varepsilon_0} \int d^3x \int d^3x^\prime
\frac{\rho_1({\bf x}) \rho_2({\bf x}^\prime)}{|{\bf x}-{\bf x}^\prime|}.
\end{equation}
Five of the six integrals are immediately completed using the property
of $\delta$-functions appearing in the charge densities
in Eqs.\,(\ref{cdenppd}) and (\ref{cdenprd}).
This leads to the interaction energy between the point dipole and the ring
\begin{equation}
E_{12} = \frac{1}{4\pi\varepsilon_0} \int_0^{2\pi} ad\phi^\prime 
\frac{\left[ ({\bf p} \cdot {\bm\lambda}) 
-3 ({\bf p} \cdot \hat{\bf r})  (\hat{\bf r} \cdot {\bm\lambda}) \right]}{r^3},
\end{equation}
where we have set
\begin{equation}
{\bf r} = {\bf x} -{\bf x}^\prime.
\end{equation}
The magnitude of ${\bf r}$ is $r$ and the unit vector
$\hat{\bf r} ={\bf r}/r$.
Here ${\bf r}$ is the position of the electric dipole
relative to the position of a line element $ad\phi^\prime$
on the ring that can be explicitly expressed using 
\begin{subequations}
\begin{eqnarray}
{\bf x} &=& \rho\cos\phi \,\hat{\bf x} 
+\rho\sin\phi \,\hat{\bf y} +(z_a+h) \,\hat{\bf z}, \\
{\bf x}^\prime &=& a\cos\phi^\prime \,\hat{\bf x} 
+a\sin\phi^\prime \,\hat{\bf y} +z_a \,\hat{\bf z},
\end{eqnarray} 
\end{subequations}
${\bf r}_0 ={\bf x}$, such that
\begin{equation}
r^2 = a^2 +h^2 +\rho^2 -2a\rho\cos(\phi-\phi^\prime).
\end{equation}

For the particular case when
the ring is polarized in the tangential direction $\hat{\bm\phi}^\prime$,
\begin{equation}
{\bm\lambda} = \lambda \hat{\bm\phi}^\prime,
\end{equation}
and the position of the point dipole is confined
on the symmetry axis of the ring,
$\rho=0$, the interaction energy is
\begin{equation}
E_{12}=0
\end{equation}
irrespective of the direction of its dipole moment ${\bf p}$.
This is so because $(\hat{\bf r}\cdot {\bm\lambda})
=\lambda(\hat{\bf r}\cdot \hat{\bm\phi}^\prime)=0$
and the integration
\begin{equation}
\int_0^{2\pi} d\phi^\prime \,({\bf p}\cdot {\bm\lambda})=0
\end{equation}
in this case, which involves integration of a cosine
function over its complete period.

For the case when the polarization of the ring is aligned to
its symmetry axis, and the electric dipole ${\bf p}$ is 
positioned on the axis and the direction of ${\bf p}$
makes an angle $\theta$ with respect to the symmetry axis,
\begin{equation}
{\bm\lambda} = \lambda \,\hat{\bf z}, 
\qquad \hat{\bm\lambda} \cdot \hat{\bf p} =\cos\theta, \qquad \rho=0.
\end{equation}
The interaction energy in this case is
\begin{equation}
E_{12} = \frac{{\bf p} \cdot {\bm\lambda}}{a^2} \frac{2\pi}{4\pi\varepsilon_0} 
\frac{a^3(a^2-2h^2)}{\left( a^2 + h^2\right)^\frac{5}{2}}.
\label{edapr}
\end{equation}
The force in the $z$-direction, a manifestation of change in energy
with respect to $h$, is
\begin{equation}
{\bf F} \cdot \hat{\bf z} 
= -\frac{\partial E_{12}}{\partial h}
= \frac{{\bf p} \cdot {\bm\lambda}}{a^3} \frac{2\pi}{4\pi\varepsilon_0} 
\frac{3a^4h(3a^2-2h^2)}{\left( a^2 + h^2\right)^\frac{7}{2}}
\end{equation}
and the torque, a manifestation of change in energy 
with respect $\theta$, is 
\begin{equation}
{\bm\tau} = -\frac{\partial E_{12}}{\partial \theta}
= \frac{{\bf p} \times {\bm\lambda}}{a^2} \frac{2\pi}{4\pi\varepsilon_0} 
\frac{a^3(a^2-2h^2)}{\left( a^2 + h^2\right)^\frac{5}{2}}.
\end{equation}
The interaction energy of Eq.\,(\ref{edapr})
is plotted as a function of $h$ for different values of orientation
angle $\theta$ in Fig.~\ref{atom-above-dielectric-ring-fig}.
The polarizations of the dipole and the ring prefer to align
anti-parallelly for $h\ll a$ and parallelly for $h\gg a$.
The energy is zero and torsion-free when $h=a/\sqrt{2}$,
represented by the point where all the curves intersect in
Fig.~\ref{atom-above-dielectric-ring-fig}.

\section{Formalism}
\label{sec-formApp}. 

We now turn our attention to the interaction between neutral
polarizable objects. As mentioned in Introduction,
the interaction energy between polarizable materials is governed by
the van der Waals and London interaction
\cite{Waals:1873sl,London:1930a,London:1930b}, which generalizes to the
Casimir and the Lifshitz interaction 
\cite{Casimir:1948pc, Lifshitz:1956sb, Dzyaloshinskii:1961fw}
when retardation effects are taken into account.
In the following, we illustrate the formalism.
This section is pedagogical and aimed towards listing
various approximations in the discussion.

The Casimir or Lifshitz interaction energy between
two dielectric materials can be expressed
as~\cite{Shajesh:2011ef}
\begin{equation}
E_{12} = \frac{\hbar c}{2} \int_{-\infty}^{\infty}
\frac{d\zeta}{2\pi}\,\text{Tr}\ln \Big[ {\bf 1} - 
{\bf \Gamma}_0\cdot {\bf T}_1 \cdot {\bf \Gamma}_0\cdot {\bf T}_2 \Big]
\label{E12=Logof}
\end{equation}
in the multiple scattering formalism, where $i\zeta=\omega/c$
is the imaginary frequency and $\omega$ is the frequency.
The free Green's dyadic ${\bf \Gamma}_0$ in Eq.~(\ref{E12=Logof})
satisfies the dyadic differential equation
\begin{equation}
-\Big[ \frac{1}{\zeta^2} {\bm\nabla} \times ({\bm\nabla} \times {\bf 1})
+ {\bf 1} \Big] \cdot {\bm\Gamma}_0 ({\bf r};i\zeta)
= {\bf 1}\, \delta^{(3)}({\bf r})
\end{equation}
with solution
\begin{equation}
{\bf \Gamma}_0({\bf r};i\zeta) = \frac{e^{-|\zeta|r}}{4\pi\,r^3}
\Big[ -u(|\zeta|r) \,{\bf 1} + v(|\zeta| r)\, \hat{\bf r} \hat{\bf r} \Big],
\label{freeGdyadic}
\end{equation}
where 
\begin{subequations}
\begin{eqnarray}
u(x) &=& 1+x+x^2, \\
v(x) &=& 3 + 3x + x^2.
\end{eqnarray}
\end{subequations}
The electric susceptibilities of the dielectric materials,
of electric permittivity ${\bm\varepsilon}_i$, is
\begin{equation}
{\bm\chi}_i({\bf x};i\zeta) 
= \frac{{\bm\varepsilon}_i({\bf x};i\zeta)}{\varepsilon_0} -{\bf 1},
\label{chi-def}
\end{equation}
where $i=1,2$, denotes respective material. In terms of
electric susceptibilities the transition matrices 
${\bf T}_i$'s in Eq.~(\ref{E12=Logof}) are
symbolically defined by the relation
\begin{equation}
{\bf T}_i = {\bm\chi}_i \cdot 
\big[{\bf 1} - {\bm\Gamma}_0 \cdot {\bm\chi}_i \big]^{-1}.
\label{Tm-chi-def}
\end{equation}
The transition matrices are themselves local quantities, but for
mathematical purposes are represented as a kernel,
\begin{equation}
{\bf T}_i({\bf x},{\bf x}^\prime;i\zeta)
= {\bf T}_i({\bf x};i\zeta) \,\delta^{(3)}({\bf x}-{\bf x}^\prime).
\end{equation}
The transition matrices are determined by solving the
Green's dyadic equation
\begin{equation}
-\Big[ \frac{1}{\zeta^2} {\bm\nabla} \times ({\bm\nabla} \times {\bf 1})
+ {\bf 1} +{\bm \chi}({\bf x};i\zeta) \Big] 
\cdot {\bm\Gamma}_i ({\bf x},{\bf x}^\prime;i\zeta)
= {\bf 1}\, \delta^{(3)}({\bf x}-{\bf x}^\prime),
\label{Gdieq}
\end{equation}
where ${\bm\Gamma}_i ({\bf x},{\bf x}^\prime;i\zeta)$ are the Green's 
dyadic for the individual materials
\begin{equation}
{\bm\Gamma}_i = {\bm\Gamma}_0 
-{\bm\Gamma}_0 \cdot {\bf T}_i \cdot {\bm\Gamma}_0.
\end{equation}


\subsection{Point atom}

Neglecting quadruple and higher moments,
the scattering matrix ${\bf T}_\text{atom}$ for an atom
with atomic dipole polarizability ${\bm\alpha}(i\zeta)$
will be modeled as 
\begin{equation}
{\bf T}_\text{atom}({\bf x};i\zeta) 
= 4\pi{\bm\alpha}(i\zeta)\, \delta^{(3)}({\bf x}-{\bf x}_0),
\label{Ti-atoms}
\end{equation}
where ${\bf x}_0$ specifies the position of the atom.
The polarizabilities are frequency dependent.
The $\delta$-functions in Eq.~(\ref{Ti-atoms}) permit a trivial evaluation
of half of the spatial integrals in the trace of Eq.~(\ref{E12=Logof}).
The frequency dependence of the atomic polarizability ${\bm\alpha}(i\zeta)$
is in general very complicated and depends on the energy levels of
the individual atom. For example,
if one assumes a two-level atom with a spherically symmetric ground state
and a spherically asymmetric excited state, then a simple model for 
the frequency dependence of the atomic polarizability is
\begin{equation}
{\bm \alpha}(i\zeta)={\bm \alpha}(0) 
\frac{\omega_0^2}{\omega_0^2+\zeta^2c^2},
\label{alpha-model}
\end{equation}
where $\omega_0$ is the excitation energy of the (two-level) atom
and ${\bm \alpha}(0)$ is its static polarizability and a measure of the
spherical asymmetry of the excited state.
In terms of the radius $a_0$ of this simple (two-level) atom,
the atomic polarizability could be further modeled as 
\begin{equation}
{\bm\alpha}(0) = \tilde{\bm\alpha}_0\, a_0^3.
\end{equation}
In this manner, one captures the content of the spherical asymmetry in the
excited state in the dimensionless parameter $\tilde {\bm\alpha}_0$.
Formally, we are ignoring the detailed structure of the transition matrix
for an atom by replacing Eq.~(\ref{Tm-chi-def}) in place
with Eq.~(\ref{Ti-atoms}), and
is valid for separation distances satisfying
\begin{equation}
|\tilde{\bm\alpha}_0|\,a_0 \ll r,
\end{equation}
because the propagators inside Eq.~(\ref{Tm-chi-def}) rarely span distances
greater than the atomic size. This approximation is valid for
dilute polarizability, $|\tilde{\bm\alpha}_0|\ll 1$, and
atomic size small compared to separation distance.
In this atomic approximation one needs to retain only the leading term
of the logarithm after expansion of Eq.~(\ref{E12=Logof}),
\begin{eqnarray}
E_\text{atom}^\text{W} ({\bm\alpha},{\bm\chi};{\bf x}_0)
= -\hbar c\int_{-\infty}^\infty d\zeta \int d^3x\,
\hspace{25mm} \nonumber \\
\,\text{tr} \, {\bm\alpha}(i\zeta) \cdot
{\bm \Gamma}({\bf x}_0-{\bf x};i\zeta) \cdot
{\bf T}({\bf x};i\zeta) \cdot
{\bm \Gamma}_0({\bf x}-{\bf x}_0;i\zeta), \hspace{5mm}
\label{E-atom-W}
\end{eqnarray}
where the trace, $\text{tr}$, is over the dyadic index, and ${\bf T}$ is
the transition matrix of the dielectric material that the atom is 
interacting with. Substituting the free Green's dyadic from 
Eq.~(\ref{freeGdyadic}) into Eq.~(\ref{E-atom-W}) we have
\begin{eqnarray}
&& E_\text{atom}^\text{W} ({\bm\alpha},{\bm\chi};{\bf x}_0) 
= -\frac{\hbar c}{16\pi^2} 
\int_{-\infty}^\infty d\zeta \int d^3x\, \frac{e^{-2|\zeta|r}}{r^6} 
\hspace{10mm} \nonumber \\ && \hspace{10mm}
\times \Big[ u^2(\zeta r)\, 
\text{tr}\{{\bm\alpha}(i\zeta)\cdot{\bf T}({\bf x};i\zeta)\}
\nonumber \\ && \hspace{14mm}
-2u(\zeta r)v(\zeta r)\, \{\hat{\bf r}\cdot{\bm\alpha}(i\zeta)\cdot
  {\bf T}({\bf x};i\zeta) \cdot\hat{\bf r}\}
\nonumber \\ && \hspace{14mm}
+ v^2(\zeta r)\, \{\hat{\bf r}\cdot{\bm\alpha}(i\zeta)\cdot\hat{\bf r}\}
\{\hat{\bf r}\cdot{\bf T}({\bf x};i\zeta)\cdot\hat{\bf r}\} \Big],
\label{E-atom-W-exp}
\end{eqnarray}
where ${\bf r}={\bf x}-{\bf x}_0$, and $r=|{\bf r}|$.
We used the property that atomic polarizability ${\bm\alpha}$
and the transition matrix ${\bf T}$ are symmetric tensors that 
let us combine the two cross terms into a single one.

\subsection{van der Waals-London approximation}

The non-retarded (van der Waals-London) regime, 
\begin{equation}
|{\bm\alpha}(0)|^{1/3} < r\ll \frac{c}{\omega_0},
\label{vdwL-approx}
\end{equation}
is a short-range approximation, where short is in  
relation to the characteristic length associated with resonant frequency.
In the regime of Eq.~(\ref{vdwL-approx}), 
the frequency dependence in Eq.~(\ref{freeGdyadic}) may be neglected
and the free dyadic in Eq.~(\ref{E-atom-W}) is approximated by the
static dipole-dipole interaction ${\bm\Gamma}_0({\bm r};0)$
to yield \cite{Craig1969ma,Craig:1969pa}, using Eq.~(\ref{E-atom-W}),
\begin{eqnarray}
&& E_\text{atom}^\text{Lon} ({\bm\alpha},{\bm\chi};{\bf x}_0)
=\hbar c\int_{-\infty}^\infty d\zeta \int d^3x 
\hspace{25mm} \nonumber \\ && \hspace{4mm}
\,\text{tr} \, {\bm\alpha}(i\zeta) \cdot
{\bm \Gamma}({\bf x}_0-{\bf x};0) \cdot
{\bf T}({\bf x};i\zeta) \cdot {\bm \Gamma}({\bf x}-{\bf x}_0;0).
\label{vdwl-gen}
\end{eqnarray}
In this the way, Eq.~(\ref{E-atom-W-exp}) simplifies to
\begin{eqnarray}
&& E_\text{atom}^\text{Lon} ({\bm\alpha},{\bm\chi};{\bf x}_0)
= -\frac{\hbar c}{16\pi^2} 
\int_{-\infty}^\infty d\zeta \int d^3x\, \frac{1}{r^6} 
\nonumber \\ && \hspace{10mm}
\times \Big[ 
\,\text{tr}\{{\bm\alpha}(i\zeta)\cdot{\bf T}({\bf x};i\zeta)\}
-6\, \{\hat{\bf r}\cdot{\bm\alpha}(i\zeta)\cdot
  {\bf T}({\bf x};i\zeta) \cdot\hat{\bf r}\}
\nonumber \\ && \hspace{14mm}
+ 9\, \{\hat{\bf r}\cdot{\bm\alpha}(i\zeta)\cdot\hat{\bf r}\}
\{\hat{\bf r}\cdot{\bf T}({\bf x};i\zeta)\cdot\hat{\bf r}\} \Big],
\label{E-atom-vdW-exp}
\end{eqnarray}
which uses $u(0)=1$ and $v(0)=3$.
For two atoms with isotropic polarizabilities, this reproduces London's
expression for the van der Waals interaction
\begin{equation}
E_\text{atom-atom}^\text{Lon} (\alpha_1{\bf 1},\alpha_2{\bf 1};r)
= -\frac{3\hbar c}{\pi\, r^6}\int_0^{\infty}
d\zeta\, \alpha_1(i\zeta)\alpha_2(i\zeta),
\label{E12WLondoniso}
\end{equation}
which is inversely proportional to the sixth power in
the separation $r$.

\subsection{Casimir-Polder approximation}
\label{CP-approx-sec}

The retarded (Casimir-Polder) regime,
\begin{equation}
|{\bm\alpha}(0)|^{1/3} <\frac{c}{\omega_0} \ll r,
\label{CP-approx}
\end{equation}
is the corresponding long-range approximation
when only fluctuations of very large time period (small frequency) 
contribute to resonance. In this approximation, the exponential dependence
on the separation distance in the free propagator of 
Eq.~(\ref{freeGdyadic}) implies that the frequency dependence of
the polarizability ${\bm \alpha_i}(i\zeta)$ in Eq.~(\ref{alpha-model})
is negligible. In the Casimir-Polder regime, the polarizabilities
can be approximated  by their static values
in Eq.~(\ref{E-atom-W}) to yield \cite{Craig1969ma,Craig:1969pa},
\begin{eqnarray}
&& E_\text{atom}^\text{CP} ({\bm\alpha},{\bm\chi};{\bf x}_0)
= -\hbar c\int_{-\infty}^\infty d\zeta \int d^3x\,
\hspace{20mm} \nonumber \\ && \hspace{5mm}
\times \,\text{tr} \, {\bm\alpha}(0) \cdot
{\bm \Gamma}({\bf x}_0-{\bf x};i\zeta) \cdot
{\bf T}({\bf x};0) \cdot {\bm \Gamma}({\bf x}-{\bf x}_0;i\zeta).
\hspace{5mm}
\label{CP-gen}
\end{eqnarray}
The $\zeta$-integration in Eq.~(\ref{E-atom-W-exp})
can be performed in this approximation to obtain
\begin{eqnarray}
&& E_\text{atom}^\text{CP} ({\bm\alpha},{\bm\chi};{\bf x}_0)
= -\frac{\hbar c}{32\pi^2} \int d^3x\, \frac{1}{r^7} 
\nonumber \\ && \hspace{5mm} 
\times \Big[ 
13\, \text{tr}\{{\bm\alpha}(0)\cdot{\bf T}({\bf x};0)\}
- 56\, \{\hat{\bf r}\cdot{\bm\alpha}(0)\cdot
  {\bf T}({\bf x};0) \cdot\hat{\bf r}\}
\nonumber \\ && \hspace{9mm} 
+ 63\, \{\hat{\bf r}\cdot{\bm\alpha}(0)\cdot\hat{\bf r}\}
\{\hat{\bf r}\cdot{\bf T}({\bf x};0)\cdot\hat{\bf r}\} \Big].
\label{E-atom-CP-exp}
\end{eqnarray}
For atoms with isotropic polarizabilities
${\bm\alpha}_1=\alpha_1{\bf 1}$ and ${\bm\alpha}_2=\alpha_2{\bf 1}$,
Eq.~(\ref{E-atom-CP-exp}) reproduces the Casimir-Polder
interaction \cite{Casmir:1947hx}
\begin{equation}
E_{12}^\text{CP}(\alpha_1{\bf 1},\alpha_2{\bf 1};r)
= -\frac{23\hbar c}{4\pi}
\frac{\alpha_1(0)\alpha_2(0)}{r^7},
\label{E12WCP33}
\end{equation}
which is inversely proportional to the seventh power in the separation $r$.

\subsection{Dilute dielectric approximation}

For a dilute dielectric material, we have
\begin{equation}
{\bm\chi} \ll {\bf 1}.
\label{dd-approx}
\end{equation}
Thus, after keeping the leading term in the expansion
of Eq.~(\ref{Tm-chi-def}),
the scattering matrix for the dilute material becomes
\begin{equation}
{\bf T}_i({\bf x},{\bf x}^\prime;i\zeta)
\sim {\bm\chi}_i({\bf x};i\zeta) \,\delta^{(3)}({\bf x}-{\bf x}^\prime).
\end{equation}
The dilute dielectric approximation of Eq.~(\ref{E-atom-W}) is 
\begin{eqnarray}
&& E_\text{atom-dd}^\text{W} = -\hbar c\int_{-\infty}^\infty d\zeta \int d^3x 
\nonumber \\ && \hspace{3mm}
\,\text{tr} \, {\bm\alpha}(i\zeta) \cdot {\bm\Gamma}_0({\bf x}_0-{\bf x};i\zeta)
\cdot {\bm\chi}({\bf x};i\zeta) \cdot {\bm\Gamma}_0({\bf x}-{\bf x}_0;i\zeta).
\hspace{7mm}
\label{E-atom-Wdd}
\end{eqnarray}
The London interaction energies in 
Eqs.~(\ref{vdwl-gen}) and (\ref{E-atom-vdW-exp}),
and the Casimir-Polder energies in
Eqs.~(\ref{CP-gen}) and (\ref{E-atom-CP-exp}),
in the approximation of Eq.~(\ref{dd-approx}) are immediately obtained
by the replacement: ${\bf T} \to {\bm\chi}$.
For example, the Casimir-Polder interaction energy between an atom
and a dilute dielectric,
using Eq.~(\ref{Ti-atoms}) and Eq.~(\ref{chi-def}), is given by
\begin{eqnarray}
&& E_\text{atom-dd}^\text{CP} ({\bm\alpha},{\bm\chi};{\bf x}_0)
= -\frac{\hbar c}{32\pi^2} \int d^3x\, \frac{1}{r^7} 
\nonumber \\ && \hspace{5mm}
\times \Big[ 
13\, \text{tr}\{{\bm\alpha}(0)\cdot{\bm\chi}({\bf x};0)\}
- 56\, \{\hat{\bf r}\cdot{\bm\alpha}(0)\cdot
  {\bm\chi}({\bf x};0) \cdot\hat{\bf r}\}
\nonumber \\ && \hspace{9mm}
+ 63\, \{\hat{\bf r}\cdot{\bm\alpha}(0)\cdot\hat{\bf r}\}
\{\hat{\bf r}\cdot{\bm\chi}({\bf x};0)\cdot\hat{\bf r}\} \Big],
\label{CP-dd-gen}
\end{eqnarray}
which is a ready-to-use expression,
because unlike the earlier approximations,
it does not require the solution for the transition matrix ${\bf T}$
obtained by solving the Green's dyadic equation in Eq.~(\ref{Gdieq}).
The dilute-dielectric approximation seems to always contain all the 
qualitative features of the interaction energy, and thus serves well
for a qualitative understanding of complicated geometries.

\section{Point atom above an infinite dielectric plate
with a circular aperture}
\label{annular-plate-section}

\begin{figure}
\includegraphics[width=9cm]{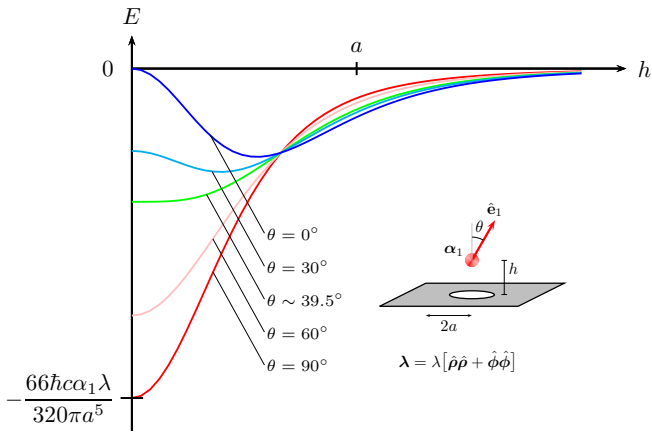}
\caption{
The energy of Eq.~(\ref{energy-atom-plate-hole})
is plotted as a function of height $h$
for an anisotropic atom above the center of a
dielectric plate with an aperture of radius $a$ as sketched in the insert.
The curves correspond to different orientation $\theta$
of the atomic polarizability.}
\label{atom-on-hole-dielectric-force-fig}
\end{figure}%

As a demonstration of the use of Eq.~(\ref{CP-dd-gen}),
in Ref.~\cite{Shajesh:2011daa}, the energy expression 
of Eq.~(\ref{CP-dd-gen}) was used to determine the interaction energy 
of an atom with polarizability,
\begin{equation}
{\bm\alpha}=\alpha_1\,\hat{\bf e}_1\, \hat{\bf e}_1,
\quad
\hat{\bf e}_1 \cdot \hat{\bf z} = \cos\theta,
\end{equation}
with a dielectric plate of infinite extent with a circular
aperture of radius $a$ with polarizability
\begin{equation}
{\bm\chi} = {\bm\lambda} \, \theta(\sqrt{x^2+y^2}-a) \, \delta(z)
\end{equation}
when the atom is positioned on the symmetry axis of the plate.

\subsection{Partially isotropic plate}

\begin{figure}
\includegraphics{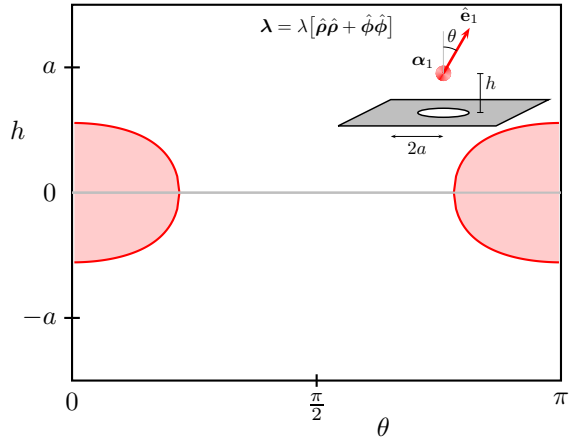}
\caption{The shaded regions in the parameter space of height $h$ and
orientation angle $\theta$ above leads to repulsion between
an anisotropic atom at a height $h$ above the center of a
dielectric plate with an aperture of radius $a$
as sketched in the insert. } 
\label{atom-on-hole-dielectric-h-vs-theta-fig}
\end{figure}

For the case when dilute dielectric plate has isotropic polarizability
in the plane of the plate we have
\begin{equation}
{\bm\lambda} = \lambda
\big[\hat{\bf x} \hat{\bf x} + \hat{\bf y} \hat{\bf y} \big] =
\lambda \big[\hat{\bm\rho} \hat{\bm\rho} + \hat{\bm\phi} \hat{\bm\phi} \big].
\end{equation}
The interaction energy has the closed form exact expression
\begin{eqnarray}
E &=& - \frac{\hbar c\alpha_1 \lambda}{64\pi} \frac{1}{5}
\frac{1}{(a^2+ h^2)^\frac{9}{2}}
\Big[ (33 a^4 +106 a^2 h^2 +38 h^4)
\nonumber \\ && \hspace{10mm}
- (33 a^4 -74 a^2 h^2 -2h^4) \cos 2\theta \Big]. \hspace{6mm}
\label{energy-atom-plate-hole}
\end{eqnarray}
Fig.~\ref{atom-on-hole-dielectric-force-fig}
shows the dependence of Eq.~(\ref{energy-atom-plate-hole}) on
the height $h$ for different orientations
$\theta$ of the anisotropic atom.
The parameter space of height $h$ and orientation angle $\theta$
that leads to repulsion has been shown as shaded regions in
Fig.~\ref{atom-on-hole-dielectric-h-vs-theta-fig}.
Repulsion ceases for orientations $\theta > \cos^{-1}(17/89)/2 \sim 39.49^\circ$.
We refer to Ref.\,\cite{Shajesh:2011daa} for a detailed discussion.
We point out that the expression for interaction energy in
Eq.\,(18) in Ref.\,\cite{Shajesh:2011daa} has errors in the 
coefficients there and should
be replaced with the correct form in Eq.\,(\ref{energy-atom-plate-hole}).
This error propagates into the analysis of repulsion there,
however, all the qualitative features of the result
and associated conclusions remain unchanged.
 
\subsection{Radially polarizable plate with aperture}

When the dilute dielectric plate is polarizable
in the radial directions, we have
\begin{equation}
{\bm\lambda} = \lambda_\rho \hat{\bm\rho} \hat{\bm\rho}.
\end{equation}
The interaction energy has the form 
\begin{eqnarray}
E &=& - \frac{\hbar c\alpha_1 \lambda_\rho}{64\pi}
\frac{1}{(a^2+ h^2)^\frac{9}{2}}
\Big[ (4 a^4 +16 a^2 h^2 +5 h^4)
\nonumber \\ && \hspace{10mm}
- (4 a^4 -20 a^2 h^2 -3h^4) \cos 2\theta \Big]. \hspace{6mm}
\label{energy-atom-plate-hole-lr}
\end{eqnarray}
The qualitative features of energy plots is completely captured
in the corresponding energy plots for the partially isotropic
case in Fig.~\ref{atom-on-hole-dielectric-h-vs-theta-fig}.
We shall return to the expression for energy
in Eq.\,(\ref{energy-atom-plate-hole-lr}) while discussing the 
case of a circular disc with circular aperture as the outer radius
of the disc extends to infinity. 

\subsection{Axially polarizable plate with aperture}

When the dilute dielectric plate is polarizable
in the direction of its symmetry axis, we have
\begin{equation}
{\bm\lambda} = \lambda_z \hat{\bf z} \hat{\bf z}.
\end{equation}
The interaction energy has the form
\begin{eqnarray}
E &=& - \frac{\hbar c\alpha_1 \lambda_z}{64\pi} \frac{1}{5}
\frac{1}{(a^2+ h^2)^\frac{9}{2}}
\Big[ (26 a^4 +17 a^2 h^2 +26 h^4)
\nonumber \\ && \hspace{10mm}
+ (26 a^4 -73 a^2 h^2 +6h^4) \cos 2\theta \Big]. \hspace{6mm}
\label{energy-atom-plate-hole-lz}
\end{eqnarray}
The energy has two orientation independent points for
$h>0$ at $h\sim 0.61a$ and $h\sim 3.44a$.
Again, we shall return to this while discussing
a disc with a circular aperture.
\begin{figure}
\includegraphics[width=9cm]{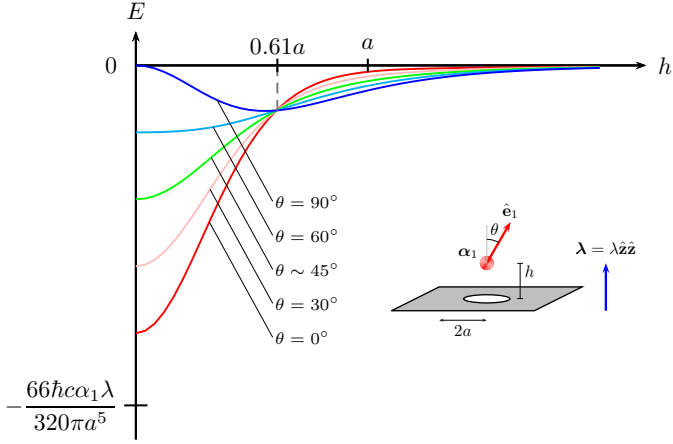}
\caption{
The energy of Eq.~(\ref{energy-atom-plate-hole-lz})
is plotted as a function of height $h$
for an anisotropic atom above the center of a
dielectric plate with an aperture of radius $a$ with
polarizability in the direction of symmetry axis of plate
as sketched in the insert.
The curves correspond to different orientation $\theta$
of the atomic polarizability.
All the curves have two orientation independent points
for $h>0$ at $h\sim 0.61a$ and $h\sim 3.44a$,
with the second orientation independent point
visible only under sufficient zoom level. }
\label{atom-on-hole-dielectric-force-e1Lz-fig}
\end{figure}%

\section{Point atom above a dielectric ring}
\label{sec-atom-ring}

The Casimir-Polder interaction energy 
between an atom and a dielectric, 
Eq.\,(\ref{Ti-atoms}) in Eq.\,(\ref{CP-dd-gen}),
is given by the expression
\begin{eqnarray}
&& E = -\frac{\hbar c}{32\pi^2} \int d^3x \frac{1}{r^7}
\Big[ 13\, \text{tr} ({\bm\alpha} \cdot {\bm\chi})
-56\, (\hat{\bf r} \cdot {\bm\alpha} \cdot {\bm\chi} \cdot \hat{\bf r})
\nonumber \\ && \hspace{34mm}
+63\, (\hat{\bf r} \cdot {\bm\alpha} \cdot \hat{\bf r})
 (\hat{\bf r} \cdot {\bm\chi} \cdot \hat{\bf r}) \Big]
\label{CPari-g}
\end{eqnarray}
where 
\begin{equation}
{\bf r} = {\bf x} - {\bf x}_0,
\end{equation}
${\bf x}_0$ being the position of the atom and 
${\bf x}$ being the integral variable scanning
the infinitesimal elements of the dielectric.
The dielectric function for the atom is described by the atomic
polarizability ${\bm\alpha}$ and the electric susceptibility 
for the interacting material is ${\bm\chi}$.
We will confine our discussion to the retarded Casimir-Polder
regime of Sec.~\ref{CP-approx-sec}, where only the static
zero frequency modes in polarizability contribute.  

\begin{figure}
\includegraphics[width=3cm]{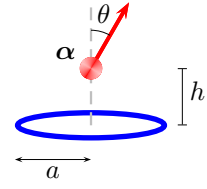}
\caption{Point atom of polarizability ${\bm\alpha}$
above a dielectric ring of polarizability ${\bm\sigma}$.
The atom is on the symmetry axis of the ring. }
\label{point-atom-above-ring-fig}
\end{figure}

Let the ring be assumed to be on the
$xy$-plane with its center at the origin. Thus,
the electric susceptibility of the ring is
\begin{equation}
{\bm\chi} = {\bm\sigma} \delta(z-0) \delta(\rho - a),
\end{equation}
where $a$ is the radius of the ring and ${\bm\sigma}$ is the
polarizability of the ring. Let the atom with polarizability
${\bm\alpha}$ be positioned on the symmetry axis of the ring
at a height $h$ above the center of the ring. Thus,
the electric susceptibility of the atom is 
\begin{equation}
{\bm\chi} = {\bm\alpha} \delta^{(3)}({\bf x}-{\bf x}_0),
\end{equation}
where
\begin{equation}
{\bf x}_0 = h\,\hat{\bf z}.
\end{equation}
See Fig.~\ref{point-atom-above-ring-fig} for an illustrative
diagram.
Since the atom is placed on the symmetry axis of the ring,
we have the simplification
\begin{equation}
{\bf r} = a\hat{\bm\rho} -h\hat{\bf z}
\end{equation}
and
\begin{equation}
r = \sqrt{a^2+h^2},
\qquad
\hat{\bf r} = \frac{a\hat{\bm\rho} -h\hat{\bf z}}
{\sqrt{a^2+h^2}}.
\label{ruv-arg}
\end{equation}
Thus, integrating over $\rho$ and $z$,
the expression for the interaction energy in Eq.\,(\ref{CPari-g})
takes the simplified form
\begin{eqnarray}
E &=& -\frac{\hbar c}{32\pi^2} 
\frac{a}{(a^2+h^2)^\frac{7}{2}} \int_0^{2\pi} d\phi
\Big[ 13\, \text{tr} ({\bm\alpha} \cdot {\bm\sigma})
\nonumber \\ && 
-56\, (\hat{\bf r} \cdot {\bm\alpha} \cdot {\bm\sigma} \cdot \hat{\bf r})
+63\, (\hat{\bf r} \cdot {\bm\alpha} \cdot \hat{\bf r})
 (\hat{\bf r} \cdot {\bm\sigma} \cdot \hat{\bf r}) \Big].
\hspace{10mm}
\label{CParj-gt}
\end{eqnarray}
We emphasize that the above expression assumes the following,
(mostly for simplification in the ensuing analysis,)
in addition to the approximations clarified  
in Section~\ref{sec-formApp}. 
\begin{enumerate}
\item We assume that the position of the point atom is exactly
on the symmetry axis of the ring, while the orientations of
the principal axes of polarizability are kept arbitrary.
This puts a severe limitation if we were interested in the
stability analysis of the atom. 
\item We assume that the polarizability tensor
of the ring is diagonal in the eigenbasis
used to the characterize the geometry of the ring.
These are chosen to be
$\hat{\bf z}$, $\hat{\bm\rho}$, and $\hat{\bm\phi}$,
with the direction of $\hat{\bf z}$ chosen along the symmetry axis
of the ring. That is, we will assume
that the polarizability tensor of the ring has the form
\begin{equation}
{\bm\sigma} = \sigma_z \,\hat{\bf z}\hat{\bf z}
+ \sigma_\rho \,\hat{\bm\rho}\hat{\bm\rho}
+ \sigma_\phi \,\hat{\bm\phi}\hat{\bm\phi}.
\end{equation}
We allow the polarizability tensor of the atom to
be completely arbitrary,  
\begin{equation}
{\bm\alpha} = \alpha_1 \hat{\bf e}_1 \hat{\bf e}_1
+ \alpha_2 \hat{\bf e}_2 \hat{\bf e}_2
+ \alpha_3 \hat{\bf e}_3 \hat{\bf e}_3.
\end{equation}
\end{enumerate}
Even with the above assumptions, the analysis involves sufficient number
of parameters required to specify the anisotropies
of the polarizations relative to the vector describing the
distance between the materials. The principal axes for the ring are
suitably chosen to be
\begin{subequations}
\begin{eqnarray}
\hat{\bm\rho} &=& \cos\phi \,\hat{\bf i} + \sin\phi\,\hat{\bf j}, \\
\hat{\bm\phi} &=& -\sin\phi \,\hat{\bf i} + \cos\phi\,\hat{\bf j}, \\
\hat{\bf z} &=& \hat{\bf k},
\end{eqnarray}
\end{subequations}
which are unit vectors associated to the cylindrical polar coordinates.
The principal axes for the atom are kept arbitrary
to allow complete generality by choosing 
\begin{subequations}
\begin{eqnarray}
\hat{\bf e}_1 &=& \hat{\bf r}_s, \\
\hat{\bf e}_2 &=& \cos\beta\, \hat{\bm\phi}_s
+ \sin\beta \,\hat{\bm\theta}_s, \\
\hat{\bf e}_3 &=& -\sin\beta\, \hat{\bm\phi}_s
+ \cos\beta \,\hat{\bm\theta}_s,
\end{eqnarray}%
\label{eieigenB-def}%
\end{subequations}%
where $\beta$ is an angle of rotation about the axes $\hat{\bf e}_1$.
Unit vectors $\hat{\bf r}_s$, $\hat{\bm \theta}_s$, and $\hat{\bm\phi}_s$
are the unit vectors associated to the spherical polar coordinates,
\begin{subequations}
\begin{eqnarray}
\hat{\bf r}_s &=& \sin\theta \cos\phi_s \,\hat{\bf i}
+ \sin\theta \sin\phi_s \,\hat{\bf j} + \cos\theta \,\hat{\bf k}, \\
\hat{\bm \theta}_s &=& \cos\theta \cos\phi_s \,\hat{\bf i}
+ \cos\theta \sin\phi_s \,\hat{\bf j} - \sin\theta \,\hat{\bf k}, \\
\hat{\bm\phi}_s &=& -\sin\phi_s \,\hat{\bf i}
+ \cos\phi_s \,\hat{\bf j}.
\end{eqnarray}
\end{subequations}
Since the axes $\hat{\bf e}_1$ need not be normal to the $x$-$y$ plane,
we choose the spherical coordinate $\phi_s$ to be different from the
cylindrical coordinate $\phi$.
The configuration of an anisotropic atom above an anisotropic ring
in terms of the above parameters involves the integration over the
angle $\phi$, which renders the interaction energy
to be dependent on the azimuth angle $\beta$ 
and independent of $\phi_s$ and $\phi$. This characteristic
allows the choice $\phi_s=0$, however, we shall refrain
and allow the algebra to bring out this feature explicitly.

\subsection{Tangentially polarizable ring}

To illustrate the anisotropic features of the interaction
in Eq.\,(\ref{CPari-g}) we begin by considering a special case.
It will turn out that this case does not permit repulsion.
Nevertheless, the associated simplification in the configuration
brings out the role of anisotropy in the interaction.

Let the polarizability of the ring be purely in the tangential
direction, that is,
\begin{equation}
{\bm\sigma} = \sigma_\phi\, \hat{\bm\phi} \hat{\bm\phi},
\end{equation}
where $\hat{\bm\phi}$ is a unit vector tangent to the ring.
Then, because this vector is always perpendicular to the
relative position vector ${\bf r}$,
\begin{equation}
\hat{\bm\phi} \cdot \hat{\bf r}=0,
\end{equation}
using Eq.\,(\ref{ruv-arg}) we conclude that two of the
three modes of interaction in Eq.\,(\ref{CParj-gt})
do not contribute. Thus, we have
\begin{eqnarray}
E &=& -\frac{\hbar c\sigma_\phi}{32\pi^2}
\frac{13a}{(a^2+h^2)^\frac{7}{2}} \int_0^{2\pi} d\phi
\, (\hat{\bm\phi} \cdot {\bm\alpha} \cdot \hat{\bm\phi}).
\label{IntE-lphiG}
\end{eqnarray}
The construction 
$(\hat{\bm\phi} \cdot {\bm\alpha} \cdot \hat{\bm\phi})$
involves projection of the polarizability of the atom
along one of the directions $\hat{\bm\phi}$ associated with the
ring, and thus is independent of the height $h$.
All the dependence in $h$ is contained in $1/(a^2+h^2)^{7/2}$.
Hence, we conclude that the energy is a monotonic
function in its dependence in height $h$. This renders the
component of the force on the atom in the direction of
$\hat{\bf z}$ to be always attractive, because it is
the negative derivative of energy with respect to $h$.
That is, a ring that is polarizable only in the tangential
direction always attracts a polarizable atom
on the symmetry axis of the ring.

This conclusion leaves this particular case
uninteresting in a discussion
on the repulsive Casimir force. However, as we mentioned earlier,
we shall carry out the discussion to a certain extent to
highlight the role of anisotropy in these interactions.

\subsubsection{\underline{Case: ${\bf\sigma}_\phi$ and $\alpha_1$}}

Let the polarizability of the atom be uniaxial
in an arbitrary direction $\hat{\bf e}_1$ given by 
\begin{equation}
{\bm\alpha} =\alpha_1\, \hat{\bf e}_1 \hat{\bf e}_1.
\end{equation}
The interaction energy of Eq.\,(\ref{IntE-lphiG})
then takes the form
\begin{eqnarray}
E &=& -\frac{\hbar c\alpha_1\sigma_\phi}{32\pi^2}
\frac{13a}{(a^2+h^2)^\frac{7}{2}} \int_0^{2\pi} d\phi
\, (\hat{\bf e}_1 \cdot \hat{\bm\phi})^2.
\label{IntE-lphiAe1}
\end{eqnarray}
Using 
\begin{equation}
\hat{\bf e}_1 \cdot \hat{\bm\phi} 
=\sin\theta \sin(\phi_s-\phi)
\end{equation}
we have
\begin{equation}
\int_0^{2\pi} d\phi \, (\hat{\bf e}_1 \cdot \hat{\bm\phi})^2
=\pi\sin^2\theta.
\end{equation}
Thus, the interaction energy in Eq.\,(\ref{IntE-lphiAe1})
takes the form
\begin{eqnarray}
E &=& - \frac{\hbar c\alpha_1\sigma_\phi}{32\pi}
\frac{13a\sin^2\theta}{(a^2+h^2)^\frac{7}{2}}.
\label{intEe1lp}
\end{eqnarray}
The orientation dependence of the polarizability 
$\hat{\bf e}_1$ with respect to the symmetry axis 
$\hat{\bf z}$ is completely contained in $\sin^2\theta$.
This is a generic feature. The interaction energy for
permanent dipoles have an orientation dependence of the
form $\cos\theta$ and that for polarizable atoms
have orientation dependence of the form 
$\cos2\theta=1-2\sin^2\theta$.
The dependence on height $h$ is monotonic,
and thus does not lead to repulsion.

\subsubsection{\underline{Case: ${\bf\sigma}_\phi$ and $\alpha_2$}}

Next, let us  assume the polarizability of the atom
to be purely in the direction $\hat{\bf e}_2$,
such that,
\begin{equation}
{\bm\alpha} =\alpha_2\, \hat{\bf e}_2 \hat{\bf e}_2.
\end{equation}
Using
\begin{equation}
\hat{\bf e}_2 \cdot \hat{\bm\phi}
=\cos\beta \cos(\phi_s-\phi)
+\cos\theta \sin\beta \sin(\phi_s-\phi)
\end{equation}
we have
\begin{equation}
\int_0^{2\pi} d\phi \, (\hat{\bf e}_2 \cdot \hat{\bm\phi})^2
=\pi(\cos^2\beta +\cos^2\theta\sin^2\beta).
\end{equation}
Thus, the interaction energy of Eq.\,(\ref{IntE-lphiG})
takes the form 
\begin{eqnarray}
E &=& - \frac{\hbar c\alpha_2\sigma_\phi}{32\pi}
\frac{13a(\cos^2\beta +\cos^2\theta\sin^2\beta)}{(a^2+h^2)^\frac{7}{2}}.
\label{intEe2lp}
\end{eqnarray}
Note that for $\theta=0$ the interaction energy has azimuthal
symmetry and is independent of $\beta$, which is expected.
Observe that for $\beta=\pi/2$ and $\theta\to\theta-(\pi/2)$
in Eq.\,(\ref{intEe2lp}), we obtain the result for
$\hat{\bf e}_1$ polarization in Eq.\,(\ref{intEe1lp}).
This corresponds to the fact that, 
in Eqs.\,(\ref{eieigenB-def}),
a rotation about $\hat{\bf e}_1$ by an angle of
$(\pi/2)-\beta$ and then a rotation about the new
$\hat{\bf e}_2$ by $-\pi/2$ takes the direction of
$\hat{\bf e}_2$ to $\hat{\bf e}_1$.
In general the interaction energy for an atom is
a linear combination of the plausible polarizabilities. 

\subsubsection{\underline{Case: ${\bf\sigma}_\phi$ and $\alpha_3$}}

Similarly, when the polarizability of the atom
is purely in the direction $\hat{\bf e}_3$, such that
\begin{equation}
{\bm\alpha} =\alpha_3\, \hat{\bf e}_3 \hat{\bf e}_3,
\end{equation}
using
\begin{equation}
\hat{\bf e}_3 \cdot \hat{\bm\phi}
=-\sin\beta \cos(\phi_s-\phi)
+\cos\theta \cos\beta \sin(\phi_s-\phi),
\end{equation}
we have
\begin{equation}
\int_0^{2\pi} d\phi \, (\hat{\bf e}_3 \cdot \hat{\bm\phi})^2
=\pi(\sin^2\beta +\cos^2\theta\cos^2\beta).
\end{equation}
Thus, the interaction energy of Eq.\,(\ref{IntE-lphiG})
takes the form
\begin{eqnarray}
E &=& - \frac{\hbar c\alpha_2\sigma_\phi}{32\pi}
\frac{13a(\sin^2\beta +\cos^2\theta\cos^2\beta)}{(a^2+h^2)^\frac{7}{2}}.
\label{intEe3lp}
\end{eqnarray}
Note that for $\theta=0$, the interaction energy has azimuthal
symmetry. Further, we observe that the interaction energy
for $\hat{\bf e}_3$ polarization in Eq.\,(\ref{intEe3lp})
is obtained from the interaction energy for $\hat{\bf e}_2$
polarization in Eq.\,(\ref{intEe2lp}) by the replacement
$\beta\to \beta +(\pi/2)$. This is true in general because
$\hat{\bf e}_2$ and $\hat{\bf e}_3$ are orthogonal vectors
in the plane perpendicular to $\hat{\bf e}_1$,
refer Eqs.\,(\ref{eieigenB-def}), and are thus related
by a ninety degree rotation in angle $\beta$.

\subsection{Radially polarizable ring}

Next, let the polarizability of the ring be purely in the radial 
direction,
\begin{equation}
{\bm\sigma} = \sigma_\rho\, \hat{\bm\rho} \hat{\bm\rho},
\end{equation}
where $\hat{\bm\rho}$ is a unit vector normal to the ring
in the $x$-$y$ plane. The expression for the interaction 
energy in Eq.\,(\ref{CParj-gt}) can be expressed in the form
\begin{eqnarray}
E &=& -\frac{\hbar c\sigma_\rho}{32\pi^2}
\frac{a}{(a^2+h^2)^\frac{7}{2}} \int_0^{2\pi} d\phi
\Big[
13 (\hat{\bm\rho} \cdot {\bm\alpha} \cdot \hat{\bm\rho})
\nonumber \\ && \hspace{5mm}
-56 (\hat{\bf r} \cdot {\bm\alpha} \cdot \hat{\bm\rho})
(\hat{\bm\rho} \cdot \hat{\bf r})
+63 (\hat{\bf r} \cdot {\bm\alpha} \cdot \hat{\bf r})
(\hat{\bm\rho} \cdot \hat{\bf r})^2
\Big]. \hspace{5mm}
\end{eqnarray}
Using the representation for $\hat{\bf r}$ in
Eq.\,(\ref{ruv-arg}) to evaluate 
$(\hat{\bm\rho} \cdot \hat{\bf r})$ we obtain
\begin{eqnarray}
E &=& -\frac{\hbar c\sigma_\rho}{32\pi^2}
\frac{a}{(a^2+h^2)^\frac{7}{2}} \int_0^{2\pi} d\phi
\bigg[ 13 (\hat{\bm\rho} \cdot {\bm\alpha} \cdot \hat{\bm\rho})
\hspace{14mm} \nonumber \\ && \hspace{15mm}
-\frac{56 a(\hat{\bf r} \cdot {\bm\alpha} \cdot \hat{\bm\rho})}
{\sqrt{a^2+h^2}}
+\frac{63 a^2 (\hat{\bf r} \cdot {\bm\alpha} \cdot \hat{\bf r})}
{(a^2+h^2)} \bigg]. \hspace{5mm}
\label{intElrG}
\end{eqnarray}

\subsubsection{\underline{Case: ${\bf\sigma}_\rho$ and $\alpha_1$}}

Let
\begin{equation}
{\bm\alpha} =\alpha_1\, \hat{\bf e}_1 \hat{\bf e}_1.
\end{equation}
Now the relevant integrals involved are
\begin{equation}
\int_0^{2\pi} d\phi \, 
(\hat{\bf r} \cdot \hat{\bf e}_1)
(\hat{\bf e}_1 \cdot \hat{\bm\rho})
=\frac{\pi a\sin^2\theta}{\sqrt{a^2+h^2}}
\end{equation}
and
\begin{equation}
\int_0^{2\pi} d\phi \, 
(\hat{\bf r} \cdot \hat{\bf e}_1)^2
=\frac{\pi (a^2\sin^2\theta +2h^2\cos^2\theta)}{(a^2+h^2)}.
\end{equation}
Using these integrals the interaction energy 
in Eq.\,(\ref{intElrG}) can be expressed in the form
\begin{eqnarray}
E &=& -\frac{\hbar c\alpha_1\sigma_\rho}{32\pi}
\frac{a}{(a^2+h^2)^\frac{11}{2}}
\Big[ 126 h^2 a^2 \cos^2\theta
\nonumber \\ && \hspace{10mm}
+(20 a^4 -30h^2 a^2 + 13 h^4) \sin^2\theta \Big]. \hspace{5mm}
\label{intE-e1lr-ex1}
\end{eqnarray}
This form for the interaction energy is useful to
compare with other polarizabilities for the atom. Another
suitable expression for interaction energy
in Eq.\,(\ref{intElrG}), obtained using half-angle formulas for
the trigonometric functions in Eq.\,(\ref{intE-e1lr-ex1}), is
\begin{eqnarray}
E &=& -\frac{\hbar c\alpha_1\sigma_\rho}{64\pi} 
\frac{a}{(a^2+h^2)^\frac{11}{2}}
\Big[ (20 a^4 + 96h^2 a^2 + 13 h^4)
\nonumber \\ && \hspace{10mm}
- (20 a^4 - 156 h^2 a^2 + 13 h^4) \cos 2\theta \Big].
\label{intE-e1lr}
\end{eqnarray}
\begin{figure}
\includegraphics[width=8cm]{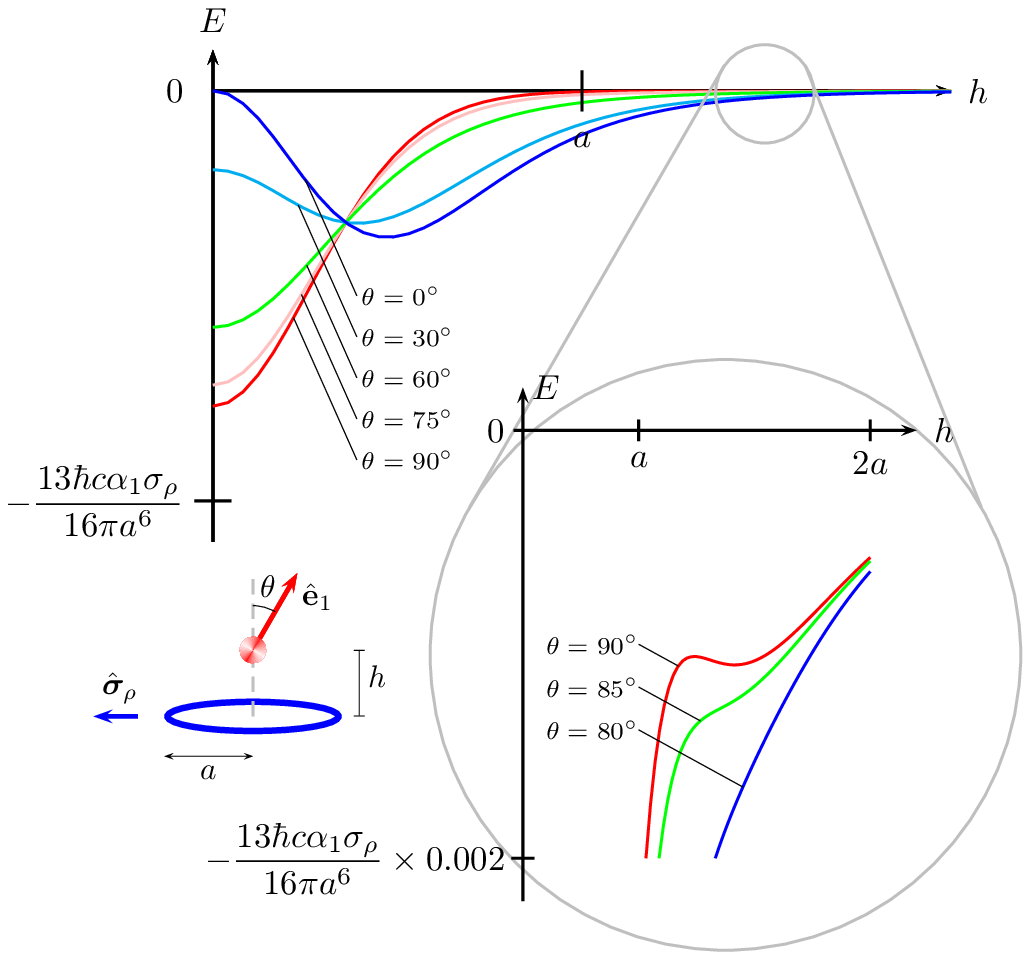}
\caption{The interaction energy between an atom with polarizability
${\bm\alpha} =\alpha_1\, \hat{\bf e}_1 \hat{\bf e}_1$
and a ring of polarizability
${\bm\sigma} = \sigma_\rho\, \hat{\bm\rho} \hat{\bm\rho}$
is illustrated in bottom left corner of figure.
The energy in Eq.\,(\ref{intE-e1lr}) is plotted as function of
height $h$ for different orientations of $\hat{\bf e}_1$
with respect to the symmetry axis of the ring.
The first repulsion occurs for $h\ll a$ when $\theta <43.43^\circ$. 
The inset provides a zoomed in view of the non-monotonicity,
second region where the switch from attraction to repulsion occurs,
which is not visible otherwise.
Repulsive force experienced by the atom in the direction of
the symmetry axis will be a manifestation of negative slopes
in these plots. }
\label{atom-above-dielectric-ringe1Lr-fig}
\end{figure}%

Figure~\ref{atom-above-dielectric-ringe1Lr-fig}
is a plot of interaction energy in Eq.\,(\ref{intE-e1lr})
as a function of height $h$ for different orientations $\theta$.
The force of repulsion is the manifestation of energy trying
to attain a minimum value, which is present whenever
the energy plots have negative slopes.
The regions with negative slope are orientation dependent,
suggesting that anisotropy in the atom's polarizability is important. 

Let us investigate the orientation dependence of
the interaction energy in Eq.\,(\ref{intE-e1lr}).
There exists a point at $h\sim 0.36a$ in 
Fig~\ref{atom-above-dielectric-ringe1Lr-fig}
where all the curves for different $\theta$ intersect.
When the domain $h$ of the plot is extended further we encounter
another such point at $h\sim 3.45a$,
which has not been captured in
Fig~\ref{atom-above-dielectric-ringe1Lr-fig}. 
These points for height $h$ are determined by the zeros of the
polynomial that is coefficient of $\cos 2\theta$
in Eq.\,(\ref{intE-e1lr}). That is,
\begin{equation}
13 h^4 - 156 h^2 a^2 + 20 a^4 =0,
\end{equation}
which have solutions
\begin{equation}
h = \pm a \sqrt{\frac{78\pm 8\sqrt{91}}{13}},
\label{hvals}
\end{equation}
or, $h=\pm 0.36\,a$ and $h=\pm 3.45\,a$.
For heights below the first torsion-free point, $h=0.36\,a$,
the atom tries to orient itself vertically ($\theta=90^\circ$),
because the energy is minimum for this orientation,
then, for heights in between the two torsion-free points,
in the range $0.36\,a <h<3.45\,a$, the atom
tries to orient itself horizontally $(\theta=0$), and
for heights beyond the second torsion-free point,
beyond $3.45\,a<h$, the atom tries to orient itself vertically. 
This analysis suggests sudden transitions between orientations
at torsion-free points without expense in energy, and could
have applications in experiments involving a beam of polarizable
atoms; for example, a beam of helium dimer. We shall
postpone discussions of such applications to another occasion.

\begin{figure}
\includegraphics{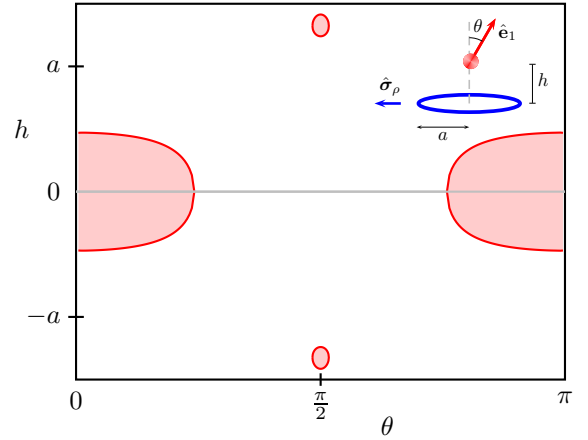}
\caption{
The shaded regions in the parameter space of height $h$ and
orientation angle $\theta$ above leads to repulsion between
an anisotropic atom with polarizability
${\bm\alpha} =\alpha_1\, \hat{\bf e}_1 \hat{\bf e}_1$
interacting with a ring of polarizability
${\bm\sigma} = \sigma_\rho\, \hat{\bm\rho} \hat{\bm\rho}$,
as illustrated in top right corner of figure
and determined by Eq.\,(\ref{regR-e1Lr}). }                             
\label{h-versus-theta-e1Lr-fig}
\end{figure}%

The negative derivative of the energy in Eq.\,(\ref{intE-e1lr})
with respect to height $h$ is the component of the force on the atom
in the vertical direction, given by
\begin{eqnarray}
F &=& -\frac{\hbar c\alpha_1\sigma_\rho}{64\pi}
\frac{ha}{(a^2+h^2)^\frac{13}{2}}
\Big[ (28 a^4 + 812h^2 a^2 + 91 h^4)
\nonumber \\ && \hspace{10mm}
- (532 a^4 - 1456 h^2 a^2 + 91 h^4) \cos 2\theta \Big].
\label{intF-e1lr}
\end{eqnarray}
For repulsion to occur the force on the atom must be positive.
So, the condition for repulsion is established using the inequality
\begin{eqnarray} 
&& (28 a^4 + 812h^2 a^2 + 91 h^4)
\nonumber \\ && \hspace{10mm}
- (532 a^4 - 1456 h^2 a^2 + 91 h^4) \cos 2\theta <0.
\end{eqnarray}
In the limit when the atom is very close to the center of
the ring, only the $h\neq 0$ terms contribute.
The atom experiences repulsion at the center of the ring
when
\begin{equation}
\cos 2\theta > \frac{28}{512},
\end{equation}
which amounts to 
$\theta <\frac{1}{2}\cos^{-1}(\frac{1}{19})\sim 43.43^\circ$.
This is consistent with plots in
Fig~\ref{atom-above-dielectric-ringe1Lr-fig}.
For general orientation $\theta$, we have repulsive regions
at height $h$ given by the inequality
\begin{eqnarray}
\frac{h}{a} < \pm \sqrt{\frac{-(812+1456\cos{2\theta})\pm 
\sqrt{D}}{182(1-\cos{2\theta})}},
\label{regR-e1Lr}
\end{eqnarray}
where
\begin{eqnarray}
D&=&(812+1456\cos{2\theta})^{2}
\nonumber \\ &&
-364(1-\cos{2\theta})(28-532\cos{2\theta}).
\label{dis-tinqeLr}
\end{eqnarray}
The solutions above are real for $D>0$. This condition
allows us to find a range of
orientations for which repulsion is possible. That is,
the discriminant $D=0$ determines a condition on
angle $\theta$ for repulsion. This is given by the inequality
obtained by simplifying Eq.\,(\ref{dis-tinqeLr}),
\begin{equation}
92+364\cos{2\theta}+273\cos^{2}{2\theta}>0.
\end{equation}
We find $86.98^\circ<\theta<93.02^\circ$
to be a range of orientations allowing repulsion.
Using $h=0$, we also find
$-43.43^\circ<\theta<43.43^\circ$ and
$136.57<\theta<223.43^\circ$
to be orientations plausible for repulsion.
In Fig.~\ref{h-versus-theta-e1Lr-fig}, we plot these
repulsively interacting regions in parameter space,
determined by Eq.\,(\ref{regR-e1Lr}).

\subsubsection{\underline{Case: ${\bf\sigma}_\rho$ and $\alpha_2$}}

Let
\begin{equation}
{\bm\alpha} =\alpha_2\, \hat{\bf e}_2 \hat{\bf e}_2.
\end{equation}
The relevant integrals involved are
\begin{equation}
\int_0^{2\pi} d\phi \,
(\hat{\bf r} \cdot \hat{\bf e}_2)
(\hat{\bf e}_2 \cdot \hat{\bm\rho})
=\frac{\pi a (\cos^2\beta +\sin^2\beta \cos^2\theta)}
{\sqrt{a^2+h^2}}
\end{equation}
and
\begin{eqnarray}
\int_0^{2\pi} d\phi \,
(\hat{\bf r} \cdot \hat{\bf e}_2)^2
&=&\frac{\pi a^2(\cos^2\beta +\sin^2\beta\cos^2\theta)} 
{(a^2+h^2)}
\nonumber \\ &&
+\frac{2\pi h^2\sin^2\beta\sin^2\theta)}{(a^2+h^2)}.
\end{eqnarray}
Using these integrals, we determine the interaction energy
in Eq.\,(\ref{intElrG}) to be 
\begin{eqnarray}
&& E = -\frac{\hbar c\alpha_2\sigma_\rho}{32\pi}
\frac{a}{(a^2+h^2)^\frac{11}{2}}
\Big[ 126 h^2 a^2 \sin^2\beta\sin^2\theta 
\hspace{10mm} \nonumber \\ && \hspace{3mm}
+ (20 a^4 -30h^2 a^2 + 13 h^4)
(\cos^2\beta +\sin^2\beta\cos^2\theta) \Big].
\label{intE-e2lr}
\end{eqnarray}
The expression for energy in Eq.\,(\ref{intE-e2lr})
has the same properties under rotations discussed
after Eq.\,(\ref{intEe2lp}). Thus, for $\beta=\pi/2$
and $\theta\to\theta-(\pi/2)$, 
the energy in Eq.\,(\ref{intE-e2lr})
becomes the energy for $\hat{\bf e}_1$ polarization
in Eq.\,(\ref{intE-e1lr-ex1}).

\subsubsection{\underline{Case: ${\bf\sigma}_\rho$ and $\alpha_3$}}

Let
\begin{equation}
{\bm\alpha} =\alpha_3\, \hat{\bf e}_3 \hat{\bf e}_3.
\end{equation}
The interaction energy for this case is given by
\begin{eqnarray}
&& E = -\frac{\hbar c\alpha_3\sigma_\rho}{32\pi}
\frac{a}{(a^2+h^2)^\frac{11}{2}} \Big[
126 h^2 a^2 \cos^2\beta\sin^2\theta
\hspace{10mm} \nonumber \\ && \hspace{3mm}
+ (20 a^4 -30h^2 a^2 + 13 h^4)
(\sin^2\beta +\cos^2\beta\cos^2\theta) \Big].
\hspace{8mm}
\label{intE-e3lr}
\end{eqnarray}
This result 
is obtained from the case of $\hat{\bf e}_2$ polarization
by a rotation of ninety degrees about the $\hat{\bf e}_3$,
which is obtained by the replacement $\beta\to (\pi/2)-\beta$
and swapping $\cos\beta \leftrightarrow \sin\beta$,
as observed after Eq.\,(\ref{intEe3lp}).

\subsection{Ring polarizable about its symmetry axis}

Finally, let the polarizability of the ring be purely
along the symmetry axis of the ring, chosen to be $\hat{\bf z}$,
\begin{equation}
{\bm\sigma} = \sigma_z\, \hat{\bf z} \hat{\bf z}.
\end{equation}
The expression for the interaction energy is of the form
\begin{eqnarray}
E &=& -\frac{\hbar c\sigma_z}{32\pi^2}
\frac{a}{(a^2+h^2)^\frac{7}{2}} \int_0^{2\pi} d\phi
\Big[
13 (\hat{\bf z} \cdot {\bm\alpha} \cdot \hat{\bf z})
\nonumber \\ && \hspace{5mm}
-56 (\hat{\bf r} \cdot {\bm\alpha} \cdot \hat{\bf z})
(\hat{\bf z} \cdot \hat{\bf r})
+63 (\hat{\bf r} \cdot {\bm\alpha} \cdot \hat{\bf r})
(\hat{\bf z} \cdot \hat{\bf r})^2
\Big]. \hspace{5mm}
\end{eqnarray}
Using Eq.\,(\ref{ruv-arg}) to evaluate 
$(\hat{\bf z} \cdot \hat{\bf r})$, we obtain
\begin{eqnarray}
E &=& -\frac{\hbar c\sigma_z}{32\pi^2}
\frac{a}{(a^2+h^2)^\frac{7}{2}} \int_0^{2\pi} d\phi
\bigg[ 13 (\hat{\bf z} \cdot {\bm\alpha} \cdot \hat{\bf z})
\hspace{14mm} \nonumber \\ && \hspace{15mm}
-\frac{56 h(\hat{\bf r} \cdot {\bm\alpha} \cdot \hat{\bf z})}
{\sqrt{a^2+h^2}}
+\frac{63 h^2 (\hat{\bf r} \cdot {\bm\alpha} \cdot \hat{\bf r})}
{(a^2+h^2)} \bigg]. \hspace{5mm}
\label{intElzG}
\end{eqnarray}

\subsubsection{\underline{Case: ${\bf\sigma}_z$ and $\alpha_1$}}
\label{sec-e1Lz}

Let
\begin{equation}
{\bm\alpha} =\alpha_1\, \hat{\bf e}_1 \hat{\bf e}_1.
\end{equation}
The relevant integrals involved are
\begin{subequations}
\begin{eqnarray}
\int_0^{2\pi} d\phi \, (\hat{\bf z} \cdot \hat{\bf e}_1)^2
&=& 2\pi \cos^2\theta, \\
\int_0^{2\pi} d\phi \, (\hat{\bf r} \cdot \hat{\bf e}_1)
(\hat{\bf e}_1 \cdot \hat{\bf z})
&=& -\frac{2\pi h \cos^2\theta}{\sqrt{a^2+h^2}}, \\
\int_0^{2\pi} d\phi \, (\hat{\bf r} \cdot \hat{\bf e}_1)^2
&=&\frac{\pi (a^2\sin^2\theta +2h^2\cos^2\theta)}
{(a^2+h^2)}. \hspace{9mm}
\end{eqnarray}
\end{subequations}
Using these integrals, the interaction energy
in Eq.\,(\ref{intElzG}) can be expressed in the form
\begin{eqnarray}
E &=& -\frac{\hbar c\alpha_1\sigma_z}{32\pi}
\frac{a}{(a^2+h^2)^\frac{11}{2}}
\Big[ 63 h^2 a^2 \sin^2\theta
\nonumber \\ && \hspace{10mm}
+(26a^4 -60h^2 a^2 + 40h^4) \cos^2\theta \Big]. \hspace{5mm}
\label{intE-e1lz-ex1}
\end{eqnarray}
The differences between the interaction energy expressions
for the radially polarizable ring in Eq.\,(\ref{intE-e1lr-ex1}) and
axially polarizable ring in Eq.\,(\ref{intE-e1lz-ex1})
are insightful and brings out the source of separate contributions
to the interaction energy.
Using half-angle formulas for
the trigonometric functions in Eq.\,(\ref{intE-e1lz-ex1}),
we can express the interaction energy in the form
\begin{eqnarray}
E &=& -\frac{\hbar c\alpha_1\sigma_z}{64\pi}
\frac{a}{(a^2+h^2)^\frac{11}{2}}
\Big[ (26 a^4 + 3h^2 a^2 + 40 h^4)
\nonumber \\ && \hspace{10mm}
+ (26 a^4 - 123 h^2 a^2 + 40 h^4) \cos 2\theta \Big].
\label{intE-e1lz}
\end{eqnarray}

\begin{figure}
\includegraphics[width=8.5cm]{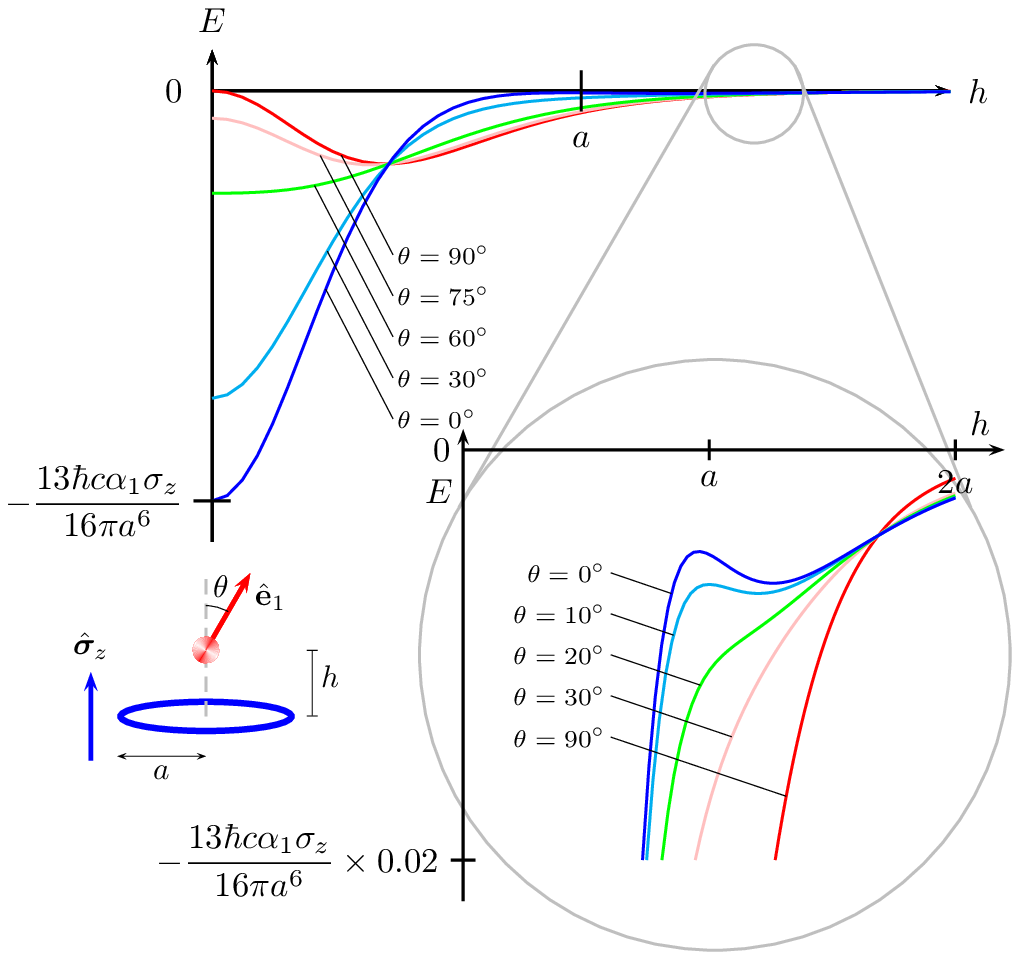}
\caption{The interaction energy of Eq.\,(\ref{intE-e1lz})
is plotted as function of height $h$ for different orientations
$\theta$. The atom has polarizability
${\bm\alpha} =\alpha_1\, \hat{\bf e}_1 \hat{\bf e}_1$
and the ring has polarizability
${\bm\sigma} = \sigma_z\, \hat{\bf z} \hat{\bf z}$,
as illustrated in bottom left corner of figure.
The inset provides a zoomed-in view of the non-monotonicity.}
\label{atom-above-dielectric-ringe1Lz-fig}
\end{figure}%

In Fig.~\ref{atom-above-dielectric-ringe1Lz-fig},
we plot the interaction energy of 
Eq.\,(\ref{intE-e1lz})
as a function of height $h$ for various orientations $\theta$.
The general features of the plot
in Fig.~\ref{atom-above-dielectric-ringe1Lz-fig}
for the axial polarization of ring
imitates that of the plot
in Fig.~\ref{atom-above-dielectric-ringe1Lr-fig}
for the radial polarization of ring.
The specifics of the interaction are different,
but the qualitative features are the same.
A striking difference is that for the radial polarization
the energy is minimized at $h=0$ for $\theta=90^\circ$,
while for the axial polarization the energy is minimized 
at $h=0$ for $\theta=0$.
That is, for both $h=0$ and $h\to\infty$,
the atom tries to align its polarization with that of the ring.

\begin{figure}
\includegraphics{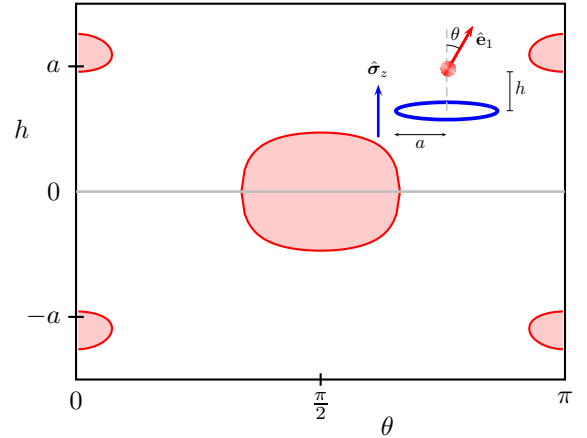}
\caption{
The shaded regions in the parameter space of height $h$ and
orientation angle $\theta$ above leads to repulsion between
an anisotropic atom with polarizability
${\bm\alpha} =\alpha_1\, \hat{\bf e}_1 \hat{\bf e}_1$
interacting with a ring of polarizability
${\bm\sigma} = \sigma_z\, \hat{\bf z} \hat{\bf z}$,
as illustrated in top right corner of figure. }
\label{h-versus-theta-e1Lz-fig}
\end{figure}%

The heights for which the interaction energy is 
orientation independent is given by the zeros of
\begin{equation}
26 a^4 - 123 h^2 a^2 + 40 h^4 =0,
\end{equation}
with solutions
\begin{equation}
h = \pm a \sqrt{\frac{123\pm \sqrt{10969}}{80}},
\end{equation}
at $h=\pm 0.48\,a$ and $h=\pm 1.69\,a$.
The expression for the component of the force along the
symmetry axis is
\begin{eqnarray}
F &=& -\frac{\hbar c\alpha_1\sigma_z}{64\pi}
\frac{7ha}{(a^2+h^2)^\frac{13}{2}}
\Big[ (40 a^4 -19h^2 a^2 +40 h^4)
\nonumber \\ && \hspace{10mm}
+ (76 a^4 -181 h^2 a^2 +40 h^4) \cos 2\theta \Big].
\label{intF-e1lz}
\end{eqnarray}
The repulsive regions in height $h$ is given by the inequality
\begin{eqnarray}
\frac{h}{a} < \pm \sqrt{\frac{(19+181\cos{2\theta})\pm
\sqrt{D}}{80(1+\cos{2\theta})}}
\end{eqnarray}
where
\begin{eqnarray}
D&=&(19+181\cos{2\theta})^{2}
\nonumber \\ &&
-160(1+\cos{2\theta})(40+76\cos{2\theta}) \nonumber \\
&=& -6039 -11682 \cos 2\theta +20601\cos^2 2\theta.
\label{dis-tinqeLz}
\end{eqnarray}
The critical angles of the orientations for which
repulsion begins is determined by $D=0$.
This shows that repulsion only occurs when
$0<\theta<13.27^\circ$ and
$166.73<\theta<180^\circ$. The other critical value
is determined by $h\to 0$, given by
\begin{equation}
\cos 2\theta = -\frac{40}{76},
\end{equation}
which has solutions $60.88^\circ<\theta<119.12^\circ$.

\subsubsection{\underline{Case: ${\bf\sigma}_z$ and $\alpha_2$}}

For
\begin{equation}
{\bm\alpha} =\alpha_2\, \hat{\bf e}_2 \hat{\bf e}_2,
\end{equation}
the interaction energy in Eq.\,(\ref{intElzG})
can be expressed in the form
\begin{eqnarray}
E &=& -\frac{\hbar c\alpha_2\sigma_z}{32\pi}
\frac{a}{(a^2+h^2)^\frac{11}{2}}
\Big[ 63 h^2 a^2 (\cos^2\beta +\sin^2\beta \cos^2\theta)
\nonumber \\ && \hspace{10mm}
+(26a^4 -60h^2 a^2 + 40h^4) \sin^2\beta \sin^2\theta \Big]. \hspace{5mm}
\label{intE-e2lz-ex1}
\end{eqnarray}
The orientation dependence in this energy should be 
compared with that for the radial polarization in
Eq.\,(\ref{intE-e2lr}).

\subsubsection{\underline{Case: ${\bf\sigma}_z$ and $\alpha_3$}}

For
\begin{equation}
{\bm\alpha} =\alpha_3\, \hat{\bf e}_3 \hat{\bf e}_3,
\end{equation}
the interaction energy in Eq.\,(\ref{intElzG}) 
can be expressed in the form
\begin{eqnarray} 
E &=& -\frac{\hbar c\alpha_3\sigma_z}{32\pi}
\frac{a}{(a^2+h^2)^\frac{11}{2}}
\Big[ 63 h^2 a^2 (\sin^2\beta +\cos^2\beta \cos^2\theta)
\nonumber \\ && \hspace{10mm}
+(26a^4 -60h^2 a^2 + 40h^4) \cos^2\beta \sin^2\theta \Big]. \hspace{5mm}
\label{intE-e3lz-ex1}
\end{eqnarray}
Again, the orientation dependence in this expression should be
contrasted with that in Eq.\,(\ref{intE-e3lr}).

\subsection{Partially isotropic ring}

If the polarizability of the ring
is isotropic in the $x$-$y$ plane we have
\begin{equation}
{\bm\sigma} = \sigma\, (\hat{\bm\rho} \hat{\bm\rho}
+\hat{\bm\phi} \hat{\bm\phi}).
\end{equation}
The interaction energy for this case will be a linear sum
of the individual energies for the radially
and tangentially polarizable cases. If the
polarizability of the atom is of the form
\begin{equation}
{\bm\alpha} =\alpha_1\, \hat{\bf e}_1 \hat{\bf e}_1,
\end{equation}
the interaction energy is obtained by adding the energies
in Eqs.\,(\ref{intEe1lp}) and (\ref{intE-e1lr}).
\begin{eqnarray}
E &=& -\frac{\hbar c\alpha_1\sigma}{64\pi}
\frac{a}{(a^2+h^2)^\frac{11}{2}}
\Big[ (33 a^4 + 122h^2 a^2 + 26 h^4)
\nonumber \\ && \hspace{10mm}
- (33 a^4 -130 h^2 a^2 + 26 h^4) \cos 2\theta \Big].
\label{intE-e1lpiso}
\end{eqnarray}
We see that the qualitative features do not change
relative to that of a radially polarizable ring.
It is of interest to inquire how the features of energy plot
change from a ring to that of a plate.
To this end, we study a polarizable annular disc.

\section{Polarizable atom above a dielectric annular disc}
\label{annular-disc-sec}

Consider a polarizable annular disc of inner radius $a$ and
outer radius $b$, with polarizability described by
\begin{equation}
{\bm\chi}={\bm\lambda} \delta(z) \theta(\rho-a) \theta(b-\rho).
\end{equation}
In the limit $b\to\infty$ this corresponds to a plate of
infinite extent with a circular aperture of radius $a$.
In the limit $a\to 0$ this corresponds to
circular disc of radius $b$.
In the simple limit $b\to a$ the annular disc is of course non existent.
However, the delicate limit of $b\to a$ in conjunction
with $\lambda\to\infty$ such that
\begin{equation}
\sigma =\lambda (b-a)
\end{equation}
is kept fixed, leads to a construction of an infinitely thin ring,
after recognizing the $\delta$-function
representation in terms of step functions
\begin{equation}
\delta(\rho-a) = \lim_{b\to a}
\frac{\theta(b-\rho) \theta(\rho-a)}{(b-a)}.
\end{equation}
In this manner, we can obtain the results for ring, and 
that for plate with circular aperture, from the results
for an annular disc.

\subsection{Partially isotropic annular disc}


Let the polarizability of the atom be of the form
\begin{equation}
{\bm\alpha} =\alpha_1\, \hat{\bf e}_1 \hat{\bf e}_1.
\end{equation}
Let the annular disc be 
isotropically polarizable in the $x$-$y$ plane,
\begin{equation}
{\bm\lambda} = \lambda {\bf 1}_\perp
=\lambda (\hat{\bf x} \hat{\bf x}
+\hat{\bf y} \hat{\bf y})
= \lambda (\hat{\bm\rho} \hat{\bm\rho}
+\hat{\bm\phi} \hat{\bm\phi}).
\end{equation}
Then, the expression for the interaction energy in Eq.\,(\ref{CPari-g})
takes the simplified form
\begin{eqnarray}
E &=& -\frac{\hbar c\alpha_1\lambda}{32\pi^2}
\int_a^b d\rho
\frac{\rho}{(\rho^2+h^2)^\frac{7}{2}} \int_0^{2\pi} d\phi
\Big[ 13 (\hat{\bf e}_1 \cdot {\bf 1}_\perp \cdot \hat{\bf e}_1)
\nonumber \\ &&
-56\, (\hat{\bf r} \cdot \hat{\bf e}_1)
(\hat{\bf e}_1 \cdot {\bf 1}_\perp \cdot \hat{\bf r})
+63\, (\hat{\bf r} \cdot \hat{\bf e}_1)^2
 (\hat{\bf r} \cdot {\bf 1}_\perp \cdot \hat{\bf r}) \Big].
\hspace{8mm}
\label{CPannD-e1li}
\end{eqnarray}
Completing the integral in $\phi$ we obtain
\begin{eqnarray}
E &=& -\frac{\hbar c\alpha_1\lambda}{64\pi}
\int_a^b d\rho \frac{\rho}{(\rho^2+h^2)^\frac{11}{2}}
\Big[ (33 \rho^4 + 122h^2 \rho^2 + 26 h^4)
\nonumber \\ && \hspace{10mm}
- (33 \rho^4 -130 h^2 \rho^2 + 26 h^4) \cos 2\theta \Big].
\label{CPannD-e1li-ap}
\end{eqnarray}
In the limit $\lambda\to\infty$ and $b\to a$ such that
$\lambda(b-a)=\sigma$ we have $\lambda\int_a^bd\rho\to\sigma$,
which then leads to the interaction energy for 
a partially isotropic ring in Eq.\,(\ref{intE-e1lpiso}).
Completing the integral in $\rho$ Eq.\,(\ref{CPannD-e1li-ap})
leads to
\begin{eqnarray}
E &=& -\frac{\hbar c\alpha_1\lambda}{320\pi}
 \frac{(-1)}{(\rho^2+h^2)^\frac{9}{2}}
\Big[ (33 \rho^4 + 106h^2 \rho^2 + 38 h^4)
\nonumber \\ && \hspace{10mm}
- (33 \rho^4 -74 h^2 \rho^2 -2 h^4) \cos 2\theta \Big]
\bigg|^{\rho=b}_{\rho=a}.
\label{CPannD-e1li-apr}
\end{eqnarray}
In the limit $b\to\infty$, this reproduces the expression for 
the energy of an infinite plate with a circular aperture
in Eq.\,(\ref{energy-atom-plate-hole}).

The characteristic features of the interaction energy
for an annular disc in Eq.\,(\ref{CPannD-e1li-apr})
are similar to those of
a partially isotropic ring in Eq.\,(\ref{intE-e1lpiso}).
These are, the two orientation independent torsion free
heights on each side of the disc, and the orientation
preferences at $h=0$ and $h\to\infty$. 
The orientation independent heights are determined by
setting the coefficient of the term with $\cos 2\theta$
to zero in Eq.\,(\ref{CPannD-e1li-apr}).
That is,
\begin{equation}
\frac{(33b^4 -74b^2h^2-2h^4)}{(b^2+h^2)^\frac{9}{2}}
-\frac{(33a^4 -74a^2h^2-2h^4)}{(a^2+h^2)^\frac{9}{2}} =0,
\end{equation}
which has four solutions for $h$. Two of the solutions are
above the disc ($h>0$), and the other two are opposite in sign
and symmetrically below the disc. Denoting the positive solutions as
$h_1$ and $h_2$, we find
\begin{subequations}
\begin{equation}
0.52 a\xleftarrow[\text{ring}]{a\leftarrow b}  
h_1 \xrightarrow[\text{plate}]{b\to\infty} 0.66 a 
\end{equation} 
and
\begin{equation}
2.18 a\xleftarrow[\text{ring}]{a\leftarrow b} 
h_2 \xrightarrow[\text{plate}]{b\to\infty} \infty,
\label{or-ind-h2-e1adap}
\end{equation}
\end{subequations}
such that $h_1$ and $h_2$ vary monotonically.
At $h=0$ the orientation preference of the atom is
to direct its polarizability in the plane of the plate ($\theta=90^\circ$).
As the atom is moved away from this position along the axis,
the orientation preference abruptly switches to
$\theta=0$ as the atom crosses the height $h=h_1$.
Afterwards it switches back to $\theta=90^\circ$ when we cross
the height $h=h_2$. 
The existence of two orientation independent heights,
on each side of disc,
thus guarantees that the orientation preference at $h=0$ will be
the same as the orientation preference at $h\to\infty$.
Conversely, since the plate ($b\to\infty$) has only one orientation
independent height, the other being at infinity,
see Eq.\,(\ref{or-ind-h2-e1adap}),
the implication is that orientation 
preference of the atom at $h=0$ is orthogonal to that
at $h\to\infty$.

\subsection{Radially polarizable annular disc}

For a polarizable atom
on the symmetry axis of an annular disc
that is polarizable radially,
\begin{equation}
{\bm\lambda} = \lambda_\rho \hat{\bm\rho} \hat{\bm\rho},
\end{equation}
the interaction energy is
\begin{eqnarray}
E &=& -\frac{\hbar c\alpha_1\lambda_\rho}{64\pi}
 \frac{(-1)}{(\rho^2+h^2)^\frac{9}{2}}
\Big[ (4 \rho^4 + 16h^2 \rho^2 + 5 h^4)
\nonumber \\ && \hspace{10mm}
- (4 \rho^4 -20 h^2 \rho^2 -3 h^4) \cos 2\theta \Big]
\bigg|^{\rho=b}_{\rho=a}.
\label{CPannD-e1lr-apr}
\end{eqnarray}
This expression tends to the expression for a ring
in Eq.\,(\ref{intE-e1lr}), and to the expression for
a plate in Eq.\,(\ref{energy-atom-plate-hole-lr}),
in the limits $b\to a$ and $b\to\infty$, respectively.

The characteristic features of the interaction energy
in Eq.\,(\ref{CPannD-e1lr-apr})
for radially polarizable annular disc qualitatively mimics
that of partially isotropically polarizable annular disc
in Eq.\,(\ref{CPannD-e1li-apr}).
The orientation independent heights are now determined by
\begin{equation}
\frac{(4b^4 -20b^2h^2-3h^4)}{(b^2+h^2)^\frac{9}{2}}
-\frac{(4a^4 -20a^2h^2-3h^4)}{(a^2+h^2)^\frac{9}{2}} =0,
\end{equation}
which again has four solutions. Two of the solutions
above the disc, ($h_1$ and $h_2$, with $0< h_1<h_2$), are
\begin{subequations}
\begin{equation}
0.36 a\xleftarrow[\text{ring}]{a\leftarrow b} 
h_1 \xrightarrow[\text{plate}]{b\to\infty} 0.44 a
\end{equation}
and
\begin{equation}
3.45 a\xleftarrow[\text{ring}]{a\leftarrow b}
h_2 \xrightarrow[\text{plate}]{b\to\infty} \infty.
\end{equation}
\end{subequations}
Again, the existence of two positive orientation independent heights 
for the case of annular disc, and for the limiting case of
disc becoming a ring,
implies that the orientation preference of the atom at
$h=0$ and $h\to\infty$ is the same.
In the limiting case of an annular disc becoming an
annular plate one of the orientation independent heights
moves to infinity, which implies that the orientation preference 
at $h=0$ is orthogonal to that at $h\to\infty$ for an annular plate.
These points have been illustrated in Fig.~\ref{E-versus-h-summary-fig}.

\begin{table}
 \begin{tabular}{ccc} 
 \hline
 Orientation &~\hspace{5mm}~& Criteria for second region of repulsion \\[0.5 ex] 
 \hline\hline
 $\theta=90^{\circ}$~~ && ~$a<b<1.257 a$  \\ 
 \hline
 $\theta=88.2^{\circ}$&& $a<b<1.201a$ \\
 \hline
 $\theta=87.3^{\circ}$ && $a<b<1.107a$\\
 \hline
\end{tabular}
\caption{This table lists the criteria when the region of repulsion
for intermediate distances can occur between
an anisotropically polarizable annular disc and
anisotropically polarizable atom for certain orientations.
The left column gives the orientation of the atom deviating
away from the symmetry axis when the disc is polarizable in
the radial direction. The right column gives the
criteria for second region of repulsion to be present.
The corresponding chart for axially polarizable annular disc
has been presented in Ref.~\cite{Marchetta:2020sap}. }
\label{table-outer-rad}
\end{table}
A motivation for analyzing the disc was to determine
when does the second region of repulsion emerge in the 
transition of the annular disc shrinking down to a ring.
These values have been evaluated numerically and
are listed in Table~\ref{table-outer-rad}.

\subsection{Axially polarizable annular disc}

When the annular disc
is polarizable along the direction of the symmetry axis
of the disc,
\begin{equation}
{\bm\lambda} = \lambda_z \hat{\bf z} \hat{\bf z},
\end{equation}
the interaction energy is
\begin{eqnarray}
E &=& -\frac{\hbar c\alpha_1\lambda_z}{64\pi} \frac{1}{5}
 \frac{(-1)}{(\rho^2+h^2)^\frac{9}{2}}
\Big[ (26 \rho^4 + 17h^2 \rho^2 + 26 h^4)
\nonumber \\ && \hspace{10mm}
+ (26 \rho^4 -73 h^2 \rho^2 +6 h^4) \cos 2\theta \Big]
\bigg|^{\rho=b}_{\rho=a}.
\label{CPannD-e1lz-apr}
\end{eqnarray}
The expression for a ring in Eq.\,(\ref{intE-e1lz}),
and the expression for a plate in Eq.\,(\ref{energy-atom-plate-hole-lz}),
are reproduced from the interaction energy of
Eq.\,(\ref{CPannD-e1lz-apr})
in the limits $b\to a$ and $b\to\infty$, respectively.

The orientation independent heights are now determined by
\begin{equation}
\frac{(26b^4 -73b^2h^2+6h^4)}{(b^2+h^2)^\frac{9}{2}}
-\frac{(26a^4 -73a^2h^2+6h^4)}{(a^2+h^2)^\frac{9}{2}} =0,
\end{equation}
with the solutions above the disc satisfying 
\begin{subequations}
\begin{equation}
0.48 a\xleftarrow[\text{ring}]{a\leftarrow b}
h_1 \xrightarrow[\text{plate}]{b\to\infty} 0.60 a
\end{equation}
and
\begin{equation}
1.69 a\xleftarrow[\text{ring}]{a\leftarrow b}
h_2 \xrightarrow[\text{plate}]{b\to\infty} 3.44a.
\end{equation}
\end{subequations}
Unlike the case of radial polarizability,
here, two orientation independent heights on each side of disc
exists even in the limit when the disc becomes a plate.
That is, the second orientation independent height
does not move away to infinity in the limiting case
of a plate.
This implies that the orientation preference of the atom at
$h=0$ and $h\to\infty$ is the same.
See illustration in Fig.~\ref{E-versus-h-summary-fig}.
This should be contrasted with that of radially
polarizable disc in which case the orientation preference
of the polarizability of the atom 
at $h=0$ is orthogonal to that at $h\to\infty$.

\section{Summary of results}
\label{sec-summary-disc}

\begin{figure}
\includegraphics[width=8.5cm]{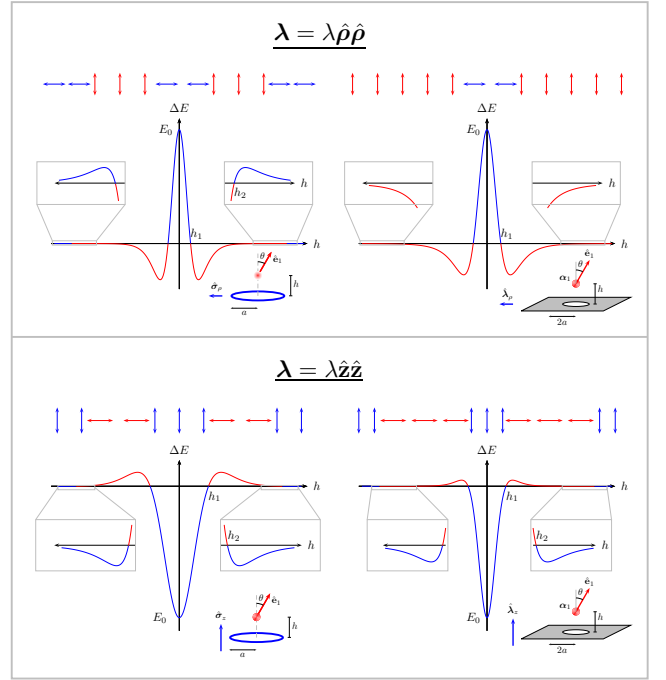}
\caption{Difference in interaction energy $\Delta E$ between
$\theta=0$ and $\theta=\pi/2$ plotted as a function of height $h$.
The intersection of the energy curves with abscissa 
represent orientation independent points, marked 
$h_1$ and $h_2$. The orientation dependence at each height is shown
as double-arrowed vectors adjacent to the horizontal axis.
Observe that the orientation preferences for the plate
is different for radial and axial polarizabilities.}
\label{E-versus-h-summary-fig}
\end{figure}%

Admittedly, we have considered too many cases and it
has been hard to keep track. 
A brief summary of our results in this paper has been presented
in Ref.~\cite{Marchetta:2020sap}.
We already pointed out that
the qualitative features of isotropically polarizability
in a plane
is identical to that of radially polarizable case.
We have also emphasized that the characteristic features of
the interaction energy are the orientation independent heights and
the orientation preferences at $h=0$ and $h=\infty$.
To this end, we feel it is sufficient to bring out the difference
between the radially polarizable case and the axially 
polarizable case.

We summarize the limiting forms of the expressions for interaction
energy for radial polarization as
\begin{subequations}
\begin{eqnarray}
{\bm\lambda} =\lambda\hat{\bm\rho} \hat{\bm\rho}: \qquad
\text{Eq.\,(\ref{CPannD-e1lr-apr})}
\begin{cases}
\xrightarrow[\text{ring}]{b\to a} 
\text{Eq.\,(\ref{intE-e1lr})} \\[7mm]
\xrightarrow[\text{plate}]{b\to\infty}
\text{Eq.\,(\ref{energy-atom-plate-hole-lr})}
\end{cases} 
\end{eqnarray}
and
\begin{eqnarray}
\text{Eq.\,(\ref{intE-e1lr})} &&
\begin{cases}
\xrightarrow{h\to 0} -E_r 20(1-\cos 2\theta), \\[3mm]
\xrightarrow{h\to\infty} 
-E_r \dfrac{a^7}{h^7} 13(1-\cos 2\theta),
\end{cases} \\[5mm]
\text{Eq.\,(\ref{energy-atom-plate-hole-lr})} &&
\begin{cases}
\xrightarrow{h\to 0} -E_p 4(1-\cos 2\theta), \\[3mm]
\xrightarrow{h\to\infty} -E_p \dfrac{a^5}{h^5} (5+3\cos 2\theta),
\end{cases}
\end{eqnarray}
\end{subequations}
where
\begin{equation}
E_r=\frac{\hbar c\alpha_1\sigma_\rho}{64\pi a^6},
\qquad
E_p=\frac{\hbar c\alpha_1\lambda_\rho}{64\pi a^5}.
\label{erep-not}
\end{equation}
The difference in the orientation dependence between the plate
and the ring arises because minimum energy for a ring
at $h\to\infty$ is decided by 
\begin{equation}
-(1-\cos 2\theta) = \begin{cases}
0, & \theta=0, \\
-2, & \theta=90^\circ,
\end{cases}
\end{equation}
and the minimum energy for a plate
at $h\to\infty$ is decided by 
\begin{equation}
-(5+3\cos 2\theta) = \begin{cases}
-8, & \theta=0, \\
-2, & \theta=90^\circ.
\end{cases}
\end{equation}
The source of this, again, can be attributed to the dyadic-dyadic
interaction.

We repeat the summary for axial polarizability as
\begin{subequations}
\begin{eqnarray}
{\bm\lambda} =\lambda\hat{\bf z} \hat{\bf z}: \qquad
\text{Eq.\,(\ref{CPannD-e1lz-apr})}
\begin{cases}
\xrightarrow[\text{ring}]{b\to a}
\text{Eq.\,(\ref{intE-e1lz})}
\\[7mm]
\xrightarrow[\text{plate}]{b\to\infty}
\text{Eq.\,(\ref{energy-atom-plate-hole-lz})}
\end{cases}
\end{eqnarray}
and
\begin{eqnarray}
\text{Eq.\,(\ref{intE-e1lz})} &&
\begin{cases}
\xrightarrow{h\to 0} -E_r 26(1+\cos 2\theta), \\[3mm]
\xrightarrow{h\to\infty} -E_r \dfrac{a^7}{h^7} 40(1+\cos 2\theta),
\end{cases} \\[5mm]
\text{Eq.\,(\ref{energy-atom-plate-hole-lz})} &&
\begin{cases}
\xrightarrow{h\to 0} -\dfrac{E_p}{5} 26(1+\cos 2\theta), \\[3mm]
\xrightarrow{h\to\infty} -\dfrac{E_p}{5} \dfrac{a^5}{h^5} (26+6\cos 2\theta),
\end{cases}
\end{eqnarray}
\end{subequations}
where $E_r$ and $E_p$ are now given by
Eq.\,(\ref{erep-not}) after swapping the subscripts
$\rho\to z$.
Identical orientation dependence between the plate
and the ring in this case arises because minimum energy for a ring
at $h\to\infty$ is decided by
\begin{equation}
-40(1+\cos 2\theta) = \begin{cases}
-80, & \theta=0, \\
0, & \theta=90^\circ,
\end{cases}
\end{equation}
and the minimum energy for a plate
at $h\to\infty$ is decided by
\begin{equation}
-(26+6\cos 2\theta) = \begin{cases}
-32, & \theta=0, \\
-20, & \theta=90^\circ.
\end{cases}
\end{equation}
The source of this can be attributed to the dyadic-dyadic
interaction, which is short of an intuitive understanding.

We plotted the energy difference between orientations at
$\theta=0^\circ$ and $\theta=90^\circ$ to highlight
the orientation independent heights and the orientation
dependence with respect to heights in Fig.~\ref{E-versus-h-summary-fig}.
Among the four plots, note that for the case of
radially polarizable plate, the atom tends to orient itself
perpendicular to the direction of polarizability of the plate
for $h\to\infty$. This contrasting feature is the key
result of our analysis.
We envision a beam of polarizable atoms
passing through an aperture in a dielectric plate
as an application that will exploit our result.

\section{Casimir machine}
\label{cas-machine-sec}

\begin{figure}
\includegraphics[width=8cm]{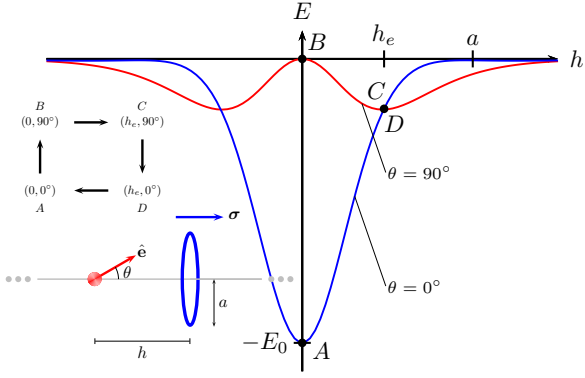}
\caption{A prototype of a Casimir machine.
Controlling the orientation $\theta$ of the polarizability of the nanoparticle
allows the movement of the polarizable ring along the symmetry axis.
For angle $\theta=0$ it is energetically favorable for the nanoparticle
to be at the center of the ring, and for $\theta=90^\circ$ it is repelled
from the ring. The plot is the interaction energy of the nanoparticle and
ring as a function of height $h$ for $\theta=0$ and $\theta=90^\circ$.
The configuration states, $A$, $B$, $C$, and $D$, in the Casimir machine
are marked on the plots and illustrated in the insets. }
\label{casimir-machine-fig}
\end{figure}

A rotaxane~\cite{Anelli:1991sma}
is a molecular structure consisting of a ring shaped 
molecule free to move on another dumbbell shaped molecule.
A rotaxane is a machine because the position of the ring
can be controlled by expending chemical or electrostatic
energy to change the associated interaction energy between
the `ring' and the `dumbbell'.
These structures have been studied with considerable interest
to find applications in constructing
molecular machines~\cite{Balzani:2008np}.
Using results from Sec.~\ref{sec-e1Lz},
we propose a prototype for a rotaxane-like machine using the
configurations studied there. An illustrative diagram of
such a machine is shown in Fig.~\ref{casimir-machine-fig}.
Casimir-Polder approximation typically require the distances to be
in the ballpark of 100\,nm, and thus our discussion here is 
applicable for plausible applications in nanotechnology.
The proposal here is superficial and presented for the
sake of captivating interest, in the sense that careful details in
the specific design has not been attended.

\begin{figure}
\includegraphics[width=7cm]{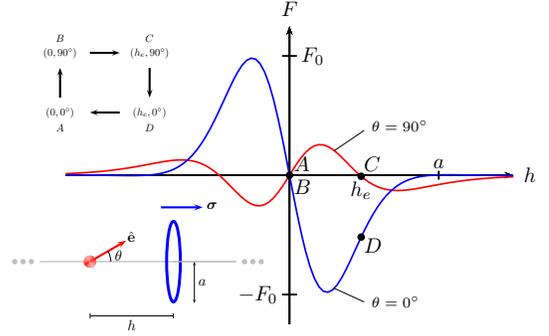}
\caption{Force verses height $h$ for $\theta=0$ and $\theta=90^\circ$.
States $A$, $B$, $C$, and $D$, for the Casimir machine are marked.
}
\label{cas-mach-force-fig}
\end{figure}

\begin{figure}
\includegraphics[width=7cm]{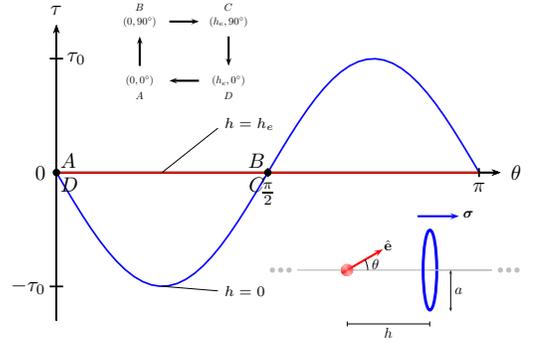}
\caption{Torque verses orientation $\theta$ for $h=0$ and $h=h_e$.
States $A$, $B$, $C$, and $D$, for the Casimir machine are marked.
}
\label{cas-mach-torque-fig}
\end{figure}

We consider a polarizable ring and a polarizable atom with 
interaction energy given by Eq.\,(\ref{intE-e1lz})
rewritten here in the form
\begin{eqnarray}
E(h,\theta) &=& -\frac{E_0}{52} \frac{a^7}{(a^2+h^2)^\frac{11}{2}}
\Big[ (26 a^4 + 3h^2 a^2 + 40 h^4)
\hspace{3mm} \nonumber \\ && \hspace{5mm}
+ (26 a^4 - 123 h^2 a^2 + 40 h^4) \cos 2\theta \Big],
\label{intE-cmac}
\end{eqnarray}
where $E_0 =13\hbar c\alpha\sigma/(16\pi a^6)$.
This energy has been plotted with respect to height $h$
in Fig.~\ref{casimir-machine-fig}
for $\theta=0$ and $\theta=90^\circ$.
Let us assume that the atom is constrained to move on
the axis. The force between the atom and ring is
given by
\begin{equation}
F =-\frac{\partial E}{\partial h}
\end{equation}
and can be expressed in the form
\begin{eqnarray}
F(h,\theta) &=& -\frac{F_0}{52} \frac{7a^8h}{(a^2+h^2)^\frac{13}{2}}
\Big[ (40 a^4 -19h^2 a^2 + 40 h^4)
\hspace{3mm} \nonumber \\ && \hspace{5mm}
+ (76 a^4 - 181 h^2 a^2 + 40 h^4) \cos 2\theta \Big],
\label{intF-cmac}
\end{eqnarray}
where $F_0 =E_0/a$.
In Fig.~\ref{cas-mach-force-fig},
we plot the force as a function of height $h$.
The torque on the atom is given by
\begin{equation}
\tau =-\frac{\partial E}{\partial\theta}
\end{equation}
and can be expressed in the form
\begin{eqnarray}
\tau(h,\theta) &=& -\frac{E_0}{26} 
\frac{a^7 (26 a^4 - 123 h^2 a^2 + 40 h^4) \sin 2\theta
}{(a^2+h^2)^\frac{11}{2}}. \hspace{7mm}
\label{intt-cmac}
\end{eqnarray}
In Fig.~\ref{cas-mach-torque-fig},
we plot the torque as a function of angle $\theta$.

\begin{figure}
\includegraphics[width=7cm]{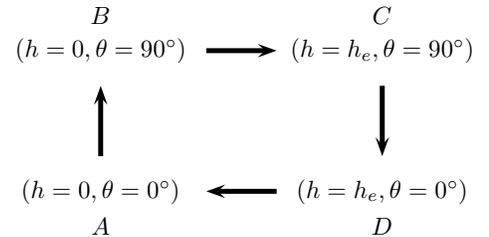}
\caption{Configuration states in the Casimir machine
in Figs.~\ref{casimir-machine-fig} to \ref{cas-mach-torque-fig}. }
\label{steps-incasmach-fig}
\end{figure}

Let us now examine the following mechanism.
Consider an initial configuration
$(h=0,\theta=0)$ denoted by state `A'
in Figs.~\ref{casimir-machine-fig} to \ref{cas-mach-torque-fig}. 
We assume that the atom can manipulate the orientation $\theta$ from 
$0^\circ$ to $90^\circ$ at the expense of internal energy.
The new state $(h=0,\theta=90^\circ)$ is denoted by state `B'
in Figs.~\ref{casimir-machine-fig} to \ref{cas-mach-torque-fig}. 
This will propel the ring away due to repulsion,
leading to the configuration
$(h=h_e,\theta=90^\circ)$ denoted by state `C'
in Figs.~\ref{casimir-machine-fig} to \ref{cas-mach-torque-fig}. 
The atom at this position manipulates the orientation $\theta$ from
$90^\circ$ to $0^\circ$ at the expense of no internal energy,
see discussion after Eq.\,(\ref{hvals}).
The new state $(h=h_e,\theta=0^\circ)$ is denoted by state `D'
in Figs.~\ref{casimir-machine-fig} to \ref{cas-mach-torque-fig}. 
This will propel the ring closer to the atom due to attraction.
The cycle $A\to B\to C\to D\to A$, 
see Fig.~\ref{steps-incasmach-fig}, thus is a machine that
is able to control the position of the ring
at the expense of internal energy in the step $A\to B$.

The change in energies in the individual steps in
the cycle are the following:
\begin{subequations}
\begin{eqnarray}
W_{A\to B} &=& -\int_{0,h=0}^\frac{\pi}{2} \tau d\theta 
= E(0,\frac{\pi}{2})-E(0,0) >0, \\
W_{B\to C} &=& -\int_{0,\theta=\frac{\pi}{2}}^{h_e} F dh
=E(h_e,\frac{\pi}{2})-E(0,\frac{\pi}{2}) <0, \hspace{10mm} \\
W_{C\to D} &=& -\int_{0,h=h_e}^\frac{\pi}{2} \tau d\theta 
=E(h_e,0)-E(h_e,\frac{\pi}{2}) =0, \\
W_{D\to A} &=& -\int_{h_e,\theta=0}^{0} F dh 
=E(0,0)-E(h_e,0) <0,
\end{eqnarray}
\end{subequations}
where the energies are evaluated using Eq.\,(\ref{intE-cmac}).
As noted, the energy input is in $A\to B$.
The step from $C\to D$ is achieved at no cost in internal energy.

\section{Conclusion and outlook}
\label{conclusion-sec}

We have derived and presented Casimir-Polder interaction energies
between an anisotropically polarizable atom and an anisotropically
polarizable ring, and when ring is replaced
with an annular disc. Then, taking the outer radius to infinity
to obtain a plate with circular aperture. 
We have identified specific configurations that
lead to repulsive Casimir-Polder forces on the atom.
We observe that when the relative orientations of the
polarizabilities of the atom and the ring are perpendicular
to each other the atom
experiences a repulsive force when it is very close to the 
center of the ring.
The energy-distance plots in 
Figs.~\ref{atom-on-hole-dielectric-force-fig},
\ref{atom-above-dielectric-ringe1Lr-fig}, and
\ref{atom-above-dielectric-ringe1Lz-fig},
mimics the energy-distance plot for an elongated needle shaped
conductor and a plate with an aperture
in Ref.~\cite{Levin:2010vo}. These plots also have the 
qualitative features of the energy-distance plot reported
for an atom and a conducting toroid
in Ref.~\cite{Abrantes2018tcp}. As suspected early on,
these interaction energies seem to be characterized by 
the interaction of the individual eigenbasis of
anisotropic polarizabilities of the objects.
These dyadic-dyadic interactions are very rich
due to the various possible interactions between the
principal polarizations.
To illustrate this richness, we have shown that
in our energy-distance plots for an atom and ring,
in addition to repulsion at short distances
when the orientations are relatively perpendicular,
we find another region of repulsion at intermediate 
distances when the orientations are relatively 
almost parallel.

We recognize that experimental observation of
repulsion resulting from anisotropy and perforated geometries
is lacking because the force is extremely weak.
Numerical estimate of these energies is of the order
of $\mu$eV, which is small. However, the energies between
conducting bodies in contrast to dielectric materials considered
in this article is expected to be larger~\cite{Venkataram:2020fpc}. 
We also note that non-monotonic Casimir forces
in the context of two interlocked corrugated geometries resembling
a `zipper' were proposed in Ref.~\cite{Rodriguez:2008gsg},
which has now been realized experimentally~\cite{Tang:2017ns}.
These developments are expected to play an
important role in the advancement of nanoscale machines.

\begin{figure}
\includegraphics[width=3cm]{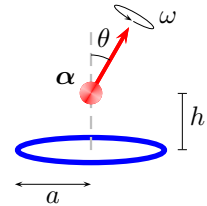}
\caption{
A spinning anisotropically polarizable atom of polarizability ${\bm\alpha}$
at a height $h$ above a dielectric ring of radius $a$.
The atom is spinning with angular speed $\omega$ about the direction
of polarization while maintaining an angle $\theta$ with the axis of the ring.
}
\label{atom-above-dielectric-ring-spin-fig}
\end{figure}%

We confined our discussion here to static configurations.
This prompts the question of whether stable equilibrium can
be achieved in these configurations when the
interactions are induced by quantum vacuum fluctuations?
By generalizing Earnshaw's
theorem to include fluctuation-induced forces,
the authors of Ref.~\cite{Rahi:2010fc} conclude that
neutral polarizable objects
cannot be in stable equilibrium in the quantum vacuum.
However, like in electrostatics,
it might still be possible to have stable equilibrium
in dynamical configurations.
We intend to study if a spinning polarizable atom,
as described in Fig.~\ref{atom-above-dielectric-ring-spin-fig},
will precess and in the process attain dynamical stable equilibrium.
This would be a Casimir-Polder analog of the
Levitron$^\text{TM}$~\cite{Berry:1996sta}.
\\

\acknowledgments

We thank Avinash Khatri, Christian Rose, Suddarsun Shivakumar,
and Preston Yun for discussions during the research.
JJM acknowledges support from the REACH Program
at Southern Illinois University--Carbondale.
PP and KVS remember Martin Schaden for collaborative assistance.

\bibliography{biblio/b20131003-casimir-top}

\end{document}